\documentclass[aip,graphicx]{revtex4-1}

\usepackage{siunitx}
\usepackage[utf8]{inputenc}
\usepackage{placeins}
\usepackage{times}
\usepackage{multirow}
\usepackage{amsmath}
\usepackage{graphicx}
\usepackage{lineno}
\usepackage[export]{adjustbox}
\usepackage{amsmath}
\usepackage{epstopdf}
\usepackage{xcolor}
\definecolor{amber}{rgb}{1.0, 0.75, 0.0}
\definecolor{brickred}{rgb}{0.7960, 0.2550, 0.3290}
\usepackage{placeins}
\usepackage{times}
\usepackage{multirow}
\usepackage{amsmath,amssymb}

\newcommand{\subfigimgtwo}[3][,]{%
  \setbox1=\hbox{\includegraphics[#1]{#3}}
  \leavevmode\rlap{\usebox1}
  \rlap{\hspace*{-5pt}\raisebox{\dimexpr\ht1-1\baselineskip}{#2}}
  \phantom{\usebox1}
}
\newcommand{\subfigimgthree}[3][,]{%
  \setbox1=\hbox{\includegraphics[#1]{#3}}
  \leavevmode\rlap{\usebox1}
  \rlap{\hspace*{-10pt}\raisebox{\dimexpr\ht1-1\baselineskip}{#2}}
  \phantom{\usebox1}
}
\newcommand{\subfigimgfour}[3][,]{%
  \setbox1=\hbox{\includegraphics[#1]{#3}}
  \leavevmode\rlap{\usebox1}
  \rlap{\hspace*{-15pt}\raisebox{\dimexpr\ht1-1\baselineskip}{#2}}
  \phantom{\usebox1}
}
\usepackage{tikz}

\usepackage{subcaption}
\usepackage{mathtools}
\usepackage{bm}

\begin{document}


\title{Influence of atmospheric conditions on the power production of utility-scale wind turbines in yaw misalignment} 



\author{Michael F. Howland}
\email[]{mhowland@stanford.edu}
\affiliation{Department of Mechanical Engineering, Stanford University, Stanford, CA 94305, U.S.A.}
\affiliation{Graduate Aerospace Laboratories (GALCIT), California Institute of Technology, Pasadena, CA 91125, U.S.A.}

\author{Carlos Moral Gonz{\'a}lez}
\affiliation{Siemens Gamesa Renewable Energy Innovation \& Technology, SL. Calle Ramírez de Arellano, 37, 28043 Madrid, Spain}
\author{Juan Jos{\'e} Pena Mart{\'i}nez}
\affiliation{Siemens Gamesa Renewable Energy Innovation \& Technology, SL. Calle Ramírez de Arellano, 37, 28043 Madrid, Spain}
\author{Jes{\'u}s Bas Quesada}
\affiliation{Siemens Gamesa Renewable Energy Innovation \& Technology, SL. Calle Ramírez de Arellano, 37, 28043 Madrid, Spain}
\author{Felipe Palou Larra\~{n}aga}
\affiliation{Siemens Gamesa Renewable Energy Innovation \& Technology, SL. Avda. Ciudad de la Innovaci{\'o}n, 2,  31621 Sarriguren, Navarra, Spain}

\author{Neeraj K. Yadav}
\affiliation{ReNew Power Private Limited, Gurugram-122009, Haryana, India}
\author{Jasvipul S. Chawla}
\affiliation{ReNew Power Private Limited, Gurugram-122009, Haryana, India}

\author{John O. Dabiri}
\email[]{jodabiri@caltech.edu}
\affiliation{Graduate Aerospace Laboratories (GALCIT), California Institute of Technology, Pasadena, CA 91125, U.S.A.}
\affiliation{Department of Mechanical and Civil Engineering, California Institute of Technology, Pasadena, CA 91125, U.S.A.}


\date{\today}

\begin{abstract}
The intentional yaw misalignment of leading, upwind turbines in a wind farm, termed wake steering, has demonstrated potential as a collective control approach for wind farm power maximization.
The optimal control strategy, and resulting effect of wake steering on wind farm power production, are in part dictated by the power degradation of the upwind yaw misaligned wind turbines.
In the atmospheric boundary layer, the wind speed and direction may vary significantly over the wind turbine rotor area, depending on atmospheric conditions and stability, resulting in freestream turbine power production which is asymmetric as a function of the direction of yaw misalignment and which varies during the diurnal cycle.
In this study, we propose a model for the power production of a wind turbine in yaw misalignment based on aerodynamic blade elements which incorporates the effects of wind speed and direction changes over the turbine rotor area in yaw misalignment.
A field experiment is performed using multiple utility-scale wind turbines to characterize the power production of yawed freestream operating turbines depending on the wind conditions, and the model is validated using the experimental data.
The resulting power production of a yaw misaligned variable speed wind turbine depends on a nonlinear interaction between the yaw misalignment, the atmospheric conditions, and the wind turbine control system.
\end{abstract}


\maketitle 

\section{Introduction}
\label{sec:intro}

Recent work has focused on the development of methodologies to increase the power production of wind farms through collective operation which considers aerodynamic interactions among individual turbines (see e.g. Kheirabadi \& Nagamune (2019) \cite{kheirabadi2019quantitative} for a recent review).
One wind farm control methodology which demonstrates potential in simulations \cite{fleming2015simulation, gebraad2016wind}, lab experiments \cite{campagnolo2016wind, bartl2018wind2}, and field experiments \cite{fleming2017field, howland2019wind, fleming2019initial, doekemeijer2020field} to increase collective turbine power production is wake steering, which entails the intentional yaw misalignment of turbines to deflect wake regions laterally away from downwind generators.
The potential for wake steering to increase wind farm power production depends on the magnitude of wake interactions between the wind turbines, the magnitude of the wake deflection as a function of yaw misalignment, and the power production lost by the yaw misaligned turbines \cite{grant1997optical}.
The power production of a wind turbine in yaw misalignment is often modeled \cite{gebraad2016wind, liew2020analytical} as
\begin{equation}
P_r=\frac{P_\gamma}{P_{\gamma_0}} \approx \cos^{P_p}(\gamma),
\label{eq:pr_general}
\end{equation}
where $P_r$ is the power ratio between the yaw misaligned $P_\gamma$ and yaw aligned $P_{\gamma_0}$ turbines.
The yaw misalignment measured at the wind turbine hub-height is given by $\gamma$.
The power ratio can also be stated in terms of the coefficient of power $C_p=P/(\frac{1}{2} \rho A u_\infty^3)$, such that $P_r = C_p(\gamma) / C_p(\gamma=0)$, where $\rho$, $A$, and $u_\infty$ are the fluid density, turbine area, and incident velocity, respectively.
Experimental wind tunnel measurements have shown that $P_p$ can vary significantly depending on the turbine model and experimental setup.
Madsen {\it et al.} (2003) \cite{madsen2003yaw} and Medici (2005) \cite{medici2005experimental} found that $P_p \approx 2$ for experimental turbine models whereas Dahlberg \& Montgomerie (2005) \cite{dahlberg2005research} found $1.88 < P_p < 5.14$ at an offshore demonstration facility.
Large eddy simulations (LES) of actuator line model wind turbines\cite{gebraad2014data} have shown $P_p\approx1.88$ for the NREL 5 MW reference turbine \cite{jonkman2009definition}.
Krogstad \& Adaramola (2012) \cite{krogstad2012performance} found that $P_p=3$ for a rotating wind turbine model in wind tunnel experiments with turbulent inflow generated by a static grid.
Bartl {\it et al.} (2018) \cite{bartl2018wind, bartl2018wind2} found that $P_p \approx 3$ for a rotating wind turbine model in wind tunnel experiments with low and high turbulence uniform inflow and sheared inflow conditions.
Schreiber {\it et al.} (2017) \cite{schreiber2017verification} and Draper {\it et al.} (2018) \cite{draper2018large} used wind tunnel experiments and LES to show that $P_p\approx 1.8$ for a wind turbine in sheared freestream conditions.
Fleming {\it et al.} (2017) \cite{fleming2017field} found $P_p\approx1.4$ for the Envision 4 MW turbine using LES and confirmed this value in a field experiment, although the number of data points beyond $|\gamma|>10^\circ$ were limited.

Wind turbine modeling methods based on blade element momentum (BEM) \cite{glauert1935airplane, madsen2003yaw} or actuator disk theory \cite{burton2011wind} both predict that $P_p=3$ (see e.g. recent discussion by Liew {\it et al.} (2020)\cite{liew2020analytical}).
Often, BEM methods leverage empirical corrections to improve the agreement with experimental data in yawed conditions (see e.g. Madsen {\it et al.} (2020) \cite{madsen2020implementation}), but these corrections are not necessarily known {\it a priori} or generally applicable.
The challenge for BEM methods to predict $P_p$, or more generally $P_r(\gamma)$ or $C_p(\gamma)$, necessitates its estimation through computationally expensive LES of wind turbine models (see discussion by Fleming {\it et al.} (2017) \cite{fleming2017field}).

Engineering wake models are often used for the selection of the optimal yaw misalignment angles for a particular wake steering scenario \cite{boersma2017tutorial}. 
Within the wake models, $P_p$ is explicitly parameterized by the user\cite{gebraad2016wind,fleming2017field,howland2019wind} or the coefficient of power $C_p$ as a function of yaw misalignment must be known {\it a priori}, which is a major barrier to wake steering deployment.
Accurate estimates of $P_p$ are required for the application of wake models for wind farm optimization since $P_p$ will dictate the trade off between the power loss at the upwind turbine against the power gain for the downwind generator.
LES studies have shown that an incorrect estimate for $P_p$ can lead to suboptimal wake steering performance \cite{howland2020optimal}.
Draper {\it et al.} (2018) \cite{draper2018large} found that the $P_p$ for a waked turbine depends on the yaw misalignment of the upwind turbine and fit experimental coefficient of power $C_p$ curves to find that $1.3<P_p<2.5$.
Liew {\it et al.} (2020) \cite{liew2020analytical} recently demonstrated in LES that $P_p=3$ is a poor estimate for wind turbines in yaw misalignment with complex, non-uniform incident wake flow and found that the value of $P_p$ depends on the incident wind conditions.
In the atmospheric boundary layer (ABL) the wind speed and direction vary as a function of height due to Coriolis, surface drag, pressure gradient, and other competing forces \cite{stull2012introduction}.
Since the value of $P_p$ will depend on the incident wind conditions, $P_p$ is not only specific to the turbine make or model but also has a functional dependence on the wind farm site and time of day, even in freestream operation.
This also presents a challenge in the comparison of literature reported values of $P_p$ with different turbine models and inflow conditions.

There have been a number of recent wake steering power maximization studies which have noted an asymmetry in the power production of a downwind turbine with respect to the direction of the yaw misalignment of the upwind turbine given full alignment \cite{fleming2015simulation, miao2016numerical, bartl2018wind2}.
Recent studies have sought to explain the noted asymmetries based on the analysis of wake dynamics.
Archer and Vasel-Be-Hagh (2019) \cite{archer2019wake} hypothesized that this asymmetry was a result of Coriolis forces which cause clockwise wake turning in the northern hemisphere \cite{van2017coriolis,howland2020coriolis}. 
Gebraad {\it et al.} (2016) \cite{gebraad2016wind} proposed that this was the result of the clockwise wind turbine blade rotation causing the wake to rotate counter-clockwise, introducing a natural rightward deflection with $\gamma=0^\circ$ and sheared, boundary layer flow.
Further, the three dimensional curled wake effect of yaw misaligned wind turbines \cite{howland2016wake, bastankhah2016experimental} may play a role in this asymmetry \cite{martinez2019aerodynamics,zong2020point} as well as the influence of wind direction changes over the turbine area on wind turbine wakes \cite{abkar2018analytical}.
However, previous studies have not considered a fundamental flow physics mechanism which would result in asymmetric thrust, angular velocity, torque, and power production for a yawed wind turbine operating in freestream conditions depending on the direction of misalignment.
As further motivation, in a recent wake steering field experiment, Doekemeijer {\it et al.} (2020) \cite{doekemeijer2020field} found an unexpected asymmetry in the $P_r$ of a freestream yaw misaligned wind turbine as a function of the sign of $\gamma$.

Aside from collective wake steering control, wind turbines attempting to minimize yaw misalignment through standard operation exhibit natural yaw offsets due to controller errors \cite{fleming2014field}, rapidly evolving wind conditions, and a trade-off between yaw error and yaw control actuation \cite{hau2013wind}.
Understanding and modeling the joint influence of yaw misalignment and the incident wind conditions on wind turbine power production is therefore useful for reducing wind farm energy production estimate error and uncertainty\cite{lackner2008uncertainty}.

The primary goal of this article is to develop a simple quantitative model which describes the power ratio $P_r(\gamma)$ as a function of wind speed and direction changes as a function of height which evolve during the diurnal cycle at a wind farm.
This model will be useful for the prediction of the power production of an arbitrary wind turbine in yaw misalignment depending on the site-specific incident wind conditions and will be directed towards controls-oriented wake modeling such as the FLORIS model \cite{bay2020floris} or lifting line model \cite{shapiro2018modelling, howland2019wind}. 
A secondary goal of this article is to perform a detailed, full-scale field experiment to characterize the power ratio $P_r$ of a wind turbine in yaw misalignment considering the broad range of realized field wind conditions for the purpose of performing a subsequent full-scale field experiment of wake steering to increase energy production.
This field experiment will also serve to validate the presented model for $P_r(\gamma)$.
The article is organized as follows: in \S \ref{sec:model_quant} the theoretical influence of the conditions occurring in stable, unstable, and approximately neutral stability states are discussed and a power ratio model is proposed.
In \S \ref{sec:exp}, the full-scale field experiment design is introduced and the experimental results and model comparisons are made in \S \ref{sec:results}.
The implication of the results on wake steering control, and wind turbine operation more broadly, are discussed in \S \ref{sec:discussion} and conclusions are given in \S \ref{sec:conclusions}.

\section{Power production model with shear and veer}
\label{sec:model_quant}

In this section, a model for the joint influence of shear, veer, and yaw misalignment on the power production of a wind turbine is proposed.
The theoretical influence of atmospheric stability on the shear and veer present in the ABL is discussed in \S \ref{sec:stability} and the power production model for general shear and veer conditions is presented in \S \ref{bem_model}.
The influence of the turbine control system is discussed in \S \ref{sec:control} and model results for canonical ABL wind profiles are given in \S \ref{sec:physics}.

\subsection{Theoretical consideration of stratification on the ABL shear and veer}
\label{sec:stability}

The magnitude of the wind direction change as a function of height depends on the atmospheric conditions.
The wind direction change over the wind turbine rotor area is defined as 
\begin{equation}
\Delta \alpha = \alpha(z=z_h+R) - \alpha(z=z_h-R),
\label{eq:dalpha}
\end{equation}
where $\alpha(z)$ is the wind direction with $0^\circ$ corresponding to north and proceeding clockwise. 
The wind turbine hub height is $z_h$.
Veer conditions result in $\Delta \alpha>0^\circ$ (e.g. flow below hub height directed to the northeast and flow above hub height directed to the east) and backing is defined as $\Delta \alpha<0^\circ$.
The wind direction change is taken as the shortest rotational path from $z_h-R$ to $z_h+R$.
The wall normal coordinate is $z$, and $x$ and $y$ are the horizontal directions.
The wind speeds in the $x$, $y$, and $z$ directions are $u$, $v$, and $w$, respectively.
The robust characterization of $\Delta \alpha$ (Eq. \ref{eq:dalpha}) relies on monotonic behavior in the wind direction $\alpha(z)$ over the wind turbine rotor area; this will be characterized in the field data in \S \ref{sec:stability_quantification}.

The wind direction change $\Delta \alpha$ depends on the effects of stratification, which is the measure of the ambient density changes in the atmosphere due to temperature and pressure variations \cite{wyngaard2010turbulence}.
With unstable stratification, convective ABL conditions present and the wind direction change as a function of height will be $\Delta \alpha \approx 0^\circ$ due to enhanced vertical mixing which reduces velocity gradients \cite{stull2012introduction}.
In the limit of neutral stratification (constant density in the atmosphere) with a balance of a geostrophic pressure gradient, Coriolis forces, and surface stress, and invoking an eddy viscosity model, the flow becomes the Ekman layer which is governed by
\begin{align}
-f_c v &= \frac{-1}{\rho} \frac{\partial P}{\partial x} + \nu_t \frac{\partial^2 u}{\partial z^2} \\
f_c u &= \frac{-1}{\rho} \frac{\partial P}{\partial y} + \nu_t \frac{\partial^2 v}{\partial z^2},
\end{align}
and a hydrostatic balance in the vertical direction where $f_c = 2 \omega \sin(\phi)$ is the Coriolis parameter, $\phi$ is the latitude, $\omega$ is the angular velocity of Earth, $P$ is the pressure, and $\nu_t$ is the eddy viscosity.
With a fixed eddy viscosity as a function of $z$, the solution is given analytically \cite{wyngaard2010turbulence} as
\begin{align}
    u &= G (1-e^{-z/\delta} \cos(z/\delta)) \\
    v &= G e^{-z/\delta} \sin(z/\delta),
\end{align}
where $G$ is the geostrophic wind speed magnitude and $\delta = \sqrt{2 \nu_t / f_c}$ is the Ekman layer depth.
The geostrophic wind is the wind associated with an exact balance between the geostrophic pressure gradient and Coriolis forces in the free atmosphere.
The Ekman layer wind direction is given by
\begin{equation}
\alpha_E(z) = \tan^{-1}\left(\frac{v(z)}{u(z)}\right) = \tan^{-1}\left(\frac{e^{-z/\delta} \sin(z/d)}{1-e^{-z/\delta} \cos(z/\delta)}\right).
\end{equation}
The wind direction variation as a function of height in this flow is termed the Ekman spiral; the wind vector turns to the left, or counter-clockwise, moving towards $z=0$, resulting in veer conditions of $\Delta \alpha >0$.
The eddy viscosity can be qualitatively modeled using a mixing length model, $l_m = \kappa z/(1+\kappa z/\lambda)$ and $\nu_t=\kappa l_m u_\tau$ with $\lambda=15$ meters, the maximum value of $l_m$ in the free atmosphere \cite{blackadar1962vertical}, and a friction velocity $u_\tau \approx 0.5$ m/s, giving $\nu_t\approx2$ m$^2$/s, a reasonable value for mid-latitudes \cite{constantin2019atmospheric}.
Overall, at a latitude of $\phi\approx25^\circ$ N, the approximate latitude of interest for the experimental wind farm, this returns a veer between the rotor diameter extent of $\Delta \alpha \approx 6^\circ$.
Further, as a result of Coriolis forces, the maximum speed in the Ekman layer occurs at a finite value of $z$ and is larger in magnitude than the geostrophic wind speed; this is termed the sub-geostrophic or low-level jet which is also present in stable ABL conditions as a result of the suppression of turbulent stresses and inertial oscillations \cite{blackadar1957boundary, van1996atmospheric}. 
Wind conditions in the atmosphere differ from the Ekman layer solution due to stratification and since the ABL is not statistically stationary (several multiples of $1/f_c$ are required for the Ekman layer flow to reach a statistically stationary state \cite{wyngaard2010turbulence, howland2018influence} during which the ABL state typically transitions).
Wind speed and direction variations as a function of height significantly modify wind farm power production through an influence on the wake recovery \cite{englberger2020rotate} and individual turbine performance \cite{murphy2019wind, sanchez2020effect}.

In stable stratification, the veering effect increases due to the suppression of turbulent production and a reduction in the boundary layer height\cite{taylor2008stratification, howland2020influence}.
Deusebio {\it et al.} (2014) \cite{deusebio2014numerical} used direct numerical simulations of stable Ekman layers and found that the veering angle generally increases with $u_\tau/Lf_c$, where 
\begin{equation}
L=-\frac{u_\tau^3 \overline{\theta_T}}{\kappa g \overline{w^\prime \theta_T^\prime}_s}
\label{eq:obhukov}
\end{equation}
is the Obhukov length with potential temperature $\theta_T$, gravity $g$, and von K{\'a}rm{\'a}n constant $\kappa$.
The surface heat flux is denoted by $\overline{w^\prime \theta_T^\prime}_s$.
As the strength of the stability increases, $L$ is positive and decreases in magnitude, and the veering angle will generally increase.
In summary, during unstable conditions which typically occur during the day, the veer will be approximately zero, while during stable conditions which occur at night, veer and a sub-geostrophic jet will present.
With increasing stable stratification, the veering angle is expected to increase.

\subsection{Blade element power ratio model}
\label{bem_model}

In yaw aligned operation with spatially uniform inflow, as a wind turbine blade rotates around its central axis of rotation, the blade angle of attack does not depend on the azimuthal position.
In yaw misalignment, sheared conditions, veered conditions, or any combination of the three, the angle of attack has a functional dependence on the azimuthal position.
Following blade element theory, we can derive a quantitative model which captures this consequence.
Kragh \& Hansen (2014) \cite{kragh2014load} developed a model for the influence of shear on the axial forces acting on a wind turbine in yaw misalignment.
Here, we use the derivation of Kragh \& Hansen \cite{kragh2014load} as a starting point and generalize the analysis to veered conditions and for power production estimation.
This analytical model is used as a starting point rather than a more complex aeroelastic solver in order to establish the first-order effects of the incident wind conditions and yaw misalignment in a complex engineering system.

Yaw alignment controllers leverage measurements of the wind direction by nacelle-mounted wind vanes in order to correct offsets between the wind direction and the nacelle position \cite{fleming2014field}.
This characterization of yaw is therefore defined as the difference between the nacelle position and the wind direction measured at hub height by the wind vane
\begin{equation}
\gamma = \alpha(z=z_h) - \beta,
\label{eq:gamma}
\end{equation}
where $\gamma$ is the yaw misalignment, $\beta$ is the nacelle position, and $\alpha(z=z_h)$ is the wind direction at hub height.
As discussed in \S \ref{sec:stability}, $\alpha(z)$ may have a functional dependence on $z$, the height above the ground, in the atmospheric boundary layer.

\begin{figure}
    \centering
    \includegraphics[width=0.9\linewidth,trim={0 0 0 0},clip]{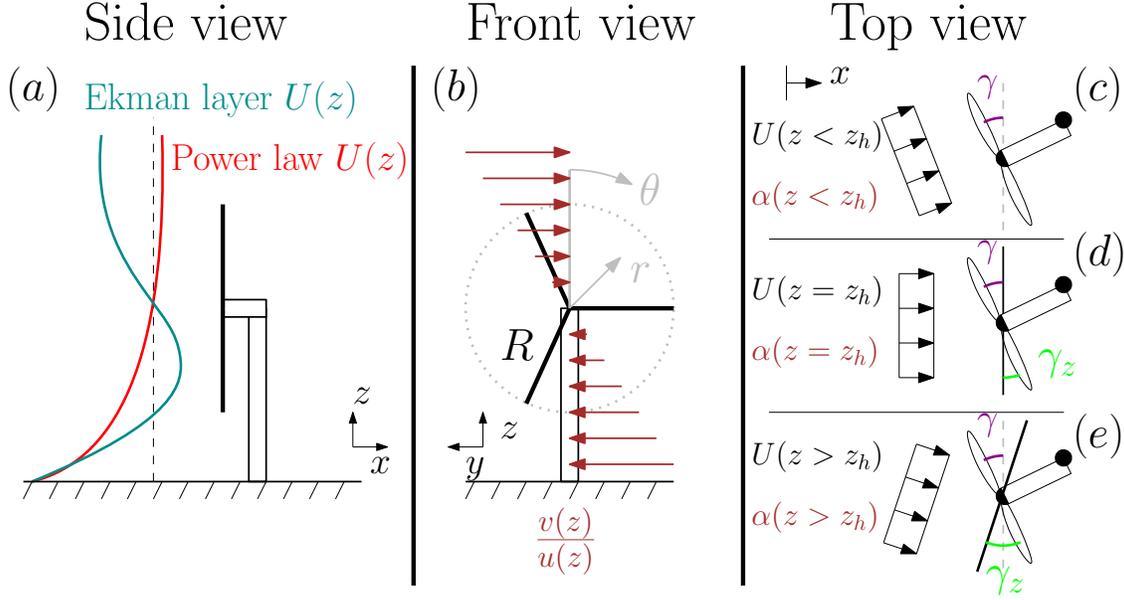}
    \caption{(a) Side view of a yaw aligned wind turbine.
    The incident wind is sheared and skewed with the wind speed $U(z)$ and direction $\alpha(z)$ depending on the height in the atmospheric boundary layer.
    Two example wind speed profiles indicative of canonical power law and Ekman layer behavior are shown.
    The Ekman layer manifests as a balance between Coriolis, pressure gradient, and surface drag forces.
    (b) Front view of a yaw aligned wind turbine with a positively veered incident inflow wind indicated by the normalized incident spanwise velocity $v(z)/u(z)$.
    The spanwise velocity profile is shown for illustrative purposes and is not generally linear.
    (c-e) Top view of a positively yaw misaligned wind turbine, which is a counter-clockwise rotation viewed from above.
    The wind turbine hub height is indicated by $z_h$.
    The top view slice is taken (c) below hub height ($z<z_h$), (d) at hub height ($z=z_h$), and (e) above hub height ($z>z_h$).
    The yaw misalignment is characterized by $\gamma=\alpha(z=z_h)-\beta$, the angle between the nacelle position $\beta$ and the hub height wind direction, $\alpha(z=z_h)$.
    The local yaw misalignment angle at the particular location of $z$ is given by $\gamma_z = \alpha(z)-\beta$.
    Given positive hub height yaw misalignment and positive veer conditions associated with Coriolis effects in the Northern Hemisphere, the wind turbine is locally more aligned below hub height ($z<z_h$) and less aligned above hub height ($z>z_h$).
    The black circle on the wind turbine nacelle is the wind speed anemometer.}
    \label{fig:turbine_schematic}
\end{figure}

\begin{figure}
    \centering
    \includegraphics[width=0.42\linewidth,trim={0 0 0 0},clip]{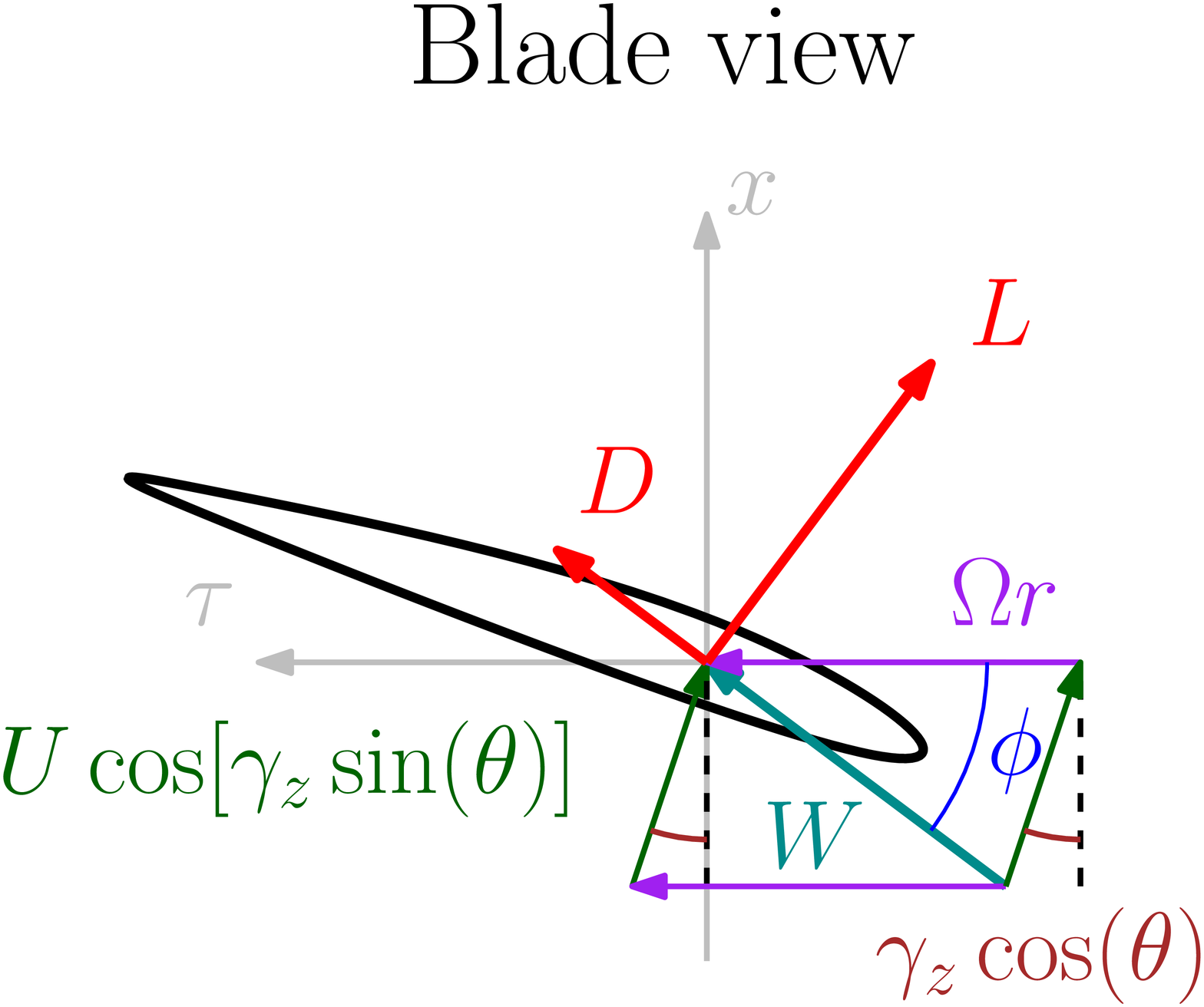}
    \caption{Blade sectional view of a positively yaw misaligned wind turbine, which is a counter-clockwise rotation viewed from above (see Figure \ref{fig:turbine_schematic}).
    The axial and tangential directions are $x$ and $\tau$, respectively.
    The turbine yaw misalignment is characterized by $\gamma=\alpha(z=z_h)-\beta$, the misalignment angle between the nacelle position $\beta$ and the hub height wind direction, $\alpha(z=z_h),$ where $z_h$ is the hub height.
    The local yaw misalignment angle incident to the blade section at the particular location of $(r,\theta)$ is given by $\gamma_z(r,\theta) = \alpha(r,\theta)-\beta$.
    The blade view shows a cross-section of a wind turbine blade as it passes through $\theta=0^\circ$.}
    \label{fig:blade_view}
\end{figure}

The wind speed incident on a blade segment is a function of its angular velocity and the incident wind velocity vector.
Neglecting the tangential induction factor (see e.g. Kragh \& Hansen (2014) \cite{kragh2014load}), the tangential velocity incident to the blade is 
\begin{equation}
u_\tau(r) = \Omega r,
\end{equation}
where $\Omega$ is the angular velocity of the blade and $r$ and $\tau$ are the radial and tangential directions, respectively.
The azimuthal angle is $\theta$.
A wind turbine's side, front, and top views with the coordinate geometry used in this study are shown in Figure \ref{fig:turbine_schematic}.
The blade sectional view and corresponding coordinate system for the blade element model is shown in Figure \ref{fig:blade_view}.

The local inflow direction over the rotor area is modified by the yaw misalignment angle in addition to wind direction variations as a function of $z$.
A local misalignment angle is defined as \begin{equation}
\gamma_z(r,\theta) = \alpha(r,\theta) - \beta.
\label{eq:gamma_local}
\end{equation}
The wind speed vector is, assuming negligible tilt, \cite{kragh2014load}
\begin{equation}
\vec{v}_{\mathrm{wind}} = \begin{bmatrix} \cos(\gamma_z \cos(\theta)) \hat{x} \\ -\sin(\gamma_z \cos(\theta)) \hat{\tau} \end{bmatrix} U \cos(\gamma_z \sin(\theta)),
\end{equation}
where $U(r,\theta)$ is the inflow wind speed corrected for axial induction effects.
The azimuthal variation of the axial induction factor is neglected in this study but could be incorporated (e.g. using Glauert's empirical correction\cite{glauert1926general}) in future work without loss of generality.
Note that the inflow wind speed $U(z)$ is transformed into polar coordinates ($r,\theta$) corresponding to the rotor plane.

The squared relative wind speed is 
\begin{equation}
W^2(r,\theta) = \left[U\cos(\gamma_z \sin(\theta)) \cos(\gamma_z \cos(\theta))\right]^2 + \left[\Omega r - U\cos(\gamma_z \sin(\theta)) \sin(\gamma_z \cos(\theta))\right]^2, 
\label{eq:w}
\end{equation}
and the inflow angle $\phi$ is
\begin{equation}
\phi = \tan^{-1}\left(\frac{U\cos(\gamma_z \sin(\theta)) \cos(\gamma_z \cos(\theta))}{\Omega r - U\cos(\gamma_z \sin(\theta)) \sin(\gamma_z \cos(\theta))}\right).
\label{eq:phi}
\end{equation}

The axial force at a particular radial section is \cite{kragh2014load} 
\begin{equation}
d f_x = \frac{1}{2} \rho c W^2 \left[C_L(\phi-\psi)\cos(\phi) + C_D(\phi-\psi)\sin(\phi)\right] dr,
\label{eq:fa}
\end{equation}
where $\psi$ incorporates blade pitch and twist at the local radial section, $c$ is the chord length, $\rho$ is the density of the incident air, and $C_L$ and $C_D$ are the lift and drag coefficients evaluated at an angle of attack of $\phi-\psi$.
The tangential force at a particular radial section is \cite{ingram2011wind}
\begin{equation}
d f_\tau = \frac{1}{2} \rho c W^2 \left[C_L(\phi-\psi)\sin(\phi) - C_D(\phi-\psi)\cos(\phi)\right] dr.
\end{equation}
The incremental torque at the particular radial section is given by
\begin{equation}
dT = r df_\tau,
\label{eq:dT}
\end{equation}
and therefore, the incremental contribution to the wind turbine power production is
\begin{equation}
dP = \Omega dT.
\label{eq:dp}
\end{equation}
Equation \ref{eq:dp} is used to compute the power ratio, defined as the power production of a yaw misaligned turbine with respect to a yaw aligned turbine
\begin{equation}
P_r = \frac{P_\gamma}{P_{\gamma_0}} =\frac{ \Omega_\gamma \int_0^{2\pi} \int_{0}^{R} c r W_\gamma^2 \left[C_L(\phi_\gamma-\psi)\sin(\phi_\gamma) - C_D(\phi_\gamma-\psi)\cos(\phi_\gamma)\right] dr d\theta}{\Omega_{\gamma_0} \int_0^{2\pi} \int_{0}^{R} c r W_{\gamma_0}^2 \left[C_L(\phi_{\gamma_0}-\psi)\sin(\phi_{\gamma_0}) - C_D(\phi_{\gamma_0}-\psi)\cos(\phi_{\gamma_0})\right] dr d\theta},
\label{eq:pr}
\end{equation}
where the subscripts of $\gamma$ and $\gamma_0$ denotes a wind turbine which is yaw misaligned or yaw aligned with respect to the hub height wind direction, respectively.
The angular velocity ratio is defined as 
\begin{equation}
\Omega_r=\frac{\Omega_\gamma}{\Omega_{\gamma_0}},
\label{eq:Or}
\end{equation}
and correspondingly, the torque ratio is given as
\begin{equation}
T_r = \frac{T_\gamma}{T_{\gamma_0}} = \frac{P_r}{\Omega_r}.
\label{eq:Tr}
\end{equation}
In order to model the power ratio, incident wind speed $U(z)$ and direction $\alpha(z)$ profiles are required, in addition to the turbine-airfoil specific coefficients of lift and drag and blade twist and pitch.
In general, for best quantitative accuracy, the lift, drag, twist, and chord tables for the specific wind turbine of study should be used if available.
For simplicity, the lift and drag coefficients, corrected for three-dimensional effects, for the NACA64 airfoil reported for the NREL 5 MW reference turbine are used \cite{jonkman2009definition} in \S \ref{sec:physics}.
The aerodynamic properties for the experimental turbine of interest are used in \S \ref{sec:results} for the field data comparisons.
Compared to aeroelastic solvers, the simple, computationally efficient model given by Eq. \ref{eq:pr} captures the leading order effects of yaw misalignment and the incident wind conditions and can be leveraged for rapid prototyping or controls-oriented optimization to predict $P_r(\gamma)$.
The proposed model does not include the assumptions associated with the momentum component of BEM theory which require empirical skewed wake corrections (see discussion by Moriarty \& Hansen (2005) \cite{moriarty2005aerodyn}).
The model will be applied to canonical ABL wind profiles in \S \ref{sec:physics} and experimentally measured wind speed $U(z)$ and direction $\alpha(z)$ profiles in \S \ref{sec:results}.

\subsubsection{Wind turbine generator torque control}
\label{sec:control}

Given velocity and wind direction profiles, the power ratio can be predicted using Eq. \ref{eq:pr} and $\Omega_r$.
The angular velocity of the blades are normalized and given by the tip-speed ratio
\begin{equation}
\lambda = \frac{\Omega R}{u_\infty}.
\end{equation}
In yaw aligned operating conditions, an optimal tip-speed ratio exists such that $C_p$ is maximized for given inflow conditions.
The angular velocity, and therefore tip-speed ratio, of a wind turbine in yaw misalignment depends on the control system in use.
Bastankhah and Port{\'e}-Agel (2017) \cite{bastankhah2017wind} found that the power ratio, and $P_p$ factor, of a wind turbine in yaw misalignment is dependent on the tip-speed ratio.
Medici (2005) \cite{medici2005experimental} used a model wind turbine embedded in a wind tunnel and found that the tip-speed ratio $\lambda_\gamma / \lambda_{\gamma_0} \sim \cos(\gamma)$ and the power ratio $P_r \sim \cos^2(\gamma)$, implying that $T_r \sim \cos(\gamma)$.
Bartl {\it et al.} (2018) \cite{bartl2018wind, bartl2018wind2} fixed the tip-speed ratio between yaw aligned and misaligned cases and found that $P_r\sim\cos^3(\gamma)$ in wind tunnel experiments.
Bastankhah and Port{\'e}-Agel (2017) \cite{bastankhah2017wind} found $P_r\sim\cos^3(\gamma)$ for a wind turbine operating at its optimal tip-speed ratio, implying that the optimal tip-speed ratio was fixed for the various yaw misalignment angles and $\Omega_r\sim\cos(\gamma)$. 
In a following study, Bastankhah and Port{\'e}-Agel (2019) \cite{bastankhah2019wind} tabulated the optimal $\Omega_\gamma$ (in rotations per minute) that returned the maximum power production as a function of the incident wind conditions and the yaw misalignment of the model turbine in wind tunnel sheared inflow.
The optimal $\Omega_\gamma$ appeared to have a weak dependence on the yaw misalignment angle (Figure 2 in Bastankhah and Port{\'e}-Agel (2019)\cite{bastankhah2019wind}).
Based on the model turbine's local wind condition measurements, the optimal set-point of $\Omega_\gamma$ was found from the lookup table and applied to the yaw misaligned turbine and $P_p\approx 2.5$, although only positive yaw angles were shown.
For these small-scale experimental model wind turbines, the operational angular velocity would be specified to Eq. \ref{eq:pr} and $P_r(\gamma)$ could be predicted based on the incident wind conditions. 

For a variable speed utility-scale horizontal axis wind turbines, the generator torque and pitch angle set-points are specified based on the wind conditions \cite[]{jonkman2009definition}.
The set-points are designed to achieve a targeted tip-speed ratio, although the steady state angular velocity, or tip-speed ratio, is a consequence of the difference between the aerodynamic and generator torque rather than an explicitly set value.
Therefore, the angular velocity achieved will depend on the generator torque setting; these two values will in turn dictate the power production.
Given a specification of $T_r$, Eq. \ref{eq:dT} can be used to compute the optimal value of $\Omega_r$ to minimize the difference between the prescribed $T_r$ and the model prediction for $T_r$.
In general, the below-rated capacity control law follows that the generator torque setting $T_c = K \Omega^2,$ where $K$ is a empirical constant which depends on the aerodynamic and electromechanical properties of the wind turbine \cite{jonkman2009definition}.
In the present study, the wind turbine generator control system is modified in yaw misalignment.
We further assume that the torque controller has reached steady-state and therefore $T_c$ is balanced exactly by the aerodynamic torque (accounting for mechanical losses and the gear-box ratio).
Equation \ref{eq:Tr} gives the aerodynamic torque as a function of the blade angular velocity and $T_c = K \Omega^2$ gives the generator controller torque as a function of the angular velocity.
Together, they provide a system of two equations and two unknowns ($\Omega_r$ and $T_r(\Omega_r)$) and can be solved with a nonlinear optimization routine (e.g. {\it fminsearch()} in Matlab \cite{lagarias1998convergence}).

\subsection{Model predictions with canonical ABL wind profiles}
\label{sec:physics}

In this section, the model proposed in \S \ref{bem_model} will be coupled with canonical ABL wind profiles to establish a qualitative physical expectation before the presentation of field results in \S \ref{sec:results}.
The wind velocity profile is approximated as a power law
\begin{equation}
u(z) = u_h (z/z_h)^{\alpha_v},
\label{eq:power_law}
\end{equation}
where $\alpha_v$ is the shear exponent and the velocity and vertical location of the wind turbine hub is given by $u_h$ and $z_h$, respectively.
While stratified ABL flows often deviate from power or logarithmic velocity profiles with the development of sub-geostrophic or low-level jets which arise from Coriolis and pressure gradient forces \cite{vera2006south}, the power law is nevertheless useful for a first order approximation \cite{stull2012introduction}.
We assume a linear profile in the wind direction as a function of height such that
\begin{equation}
\frac{d\alpha(z)}{dz}=\frac{\alpha(z_h+R)-\alpha(z_h-R)}{(z_h+R)-(z_h-R)},
\label{eq:dalpha_linear}
\end{equation}
where $\alpha(z_h+R)$ and $\alpha(z_h-R)$ are prescribed to give the veer over the turbine face.

\begin{figure}
  \centering
  \begin{tabular}{@{}p{0.33\linewidth}@{\quad}p{0.33\linewidth}@{\quad}p{0.33\linewidth}@{}}
  \subfigimgtwo[width=\linewidth,valign=t]{(a)}{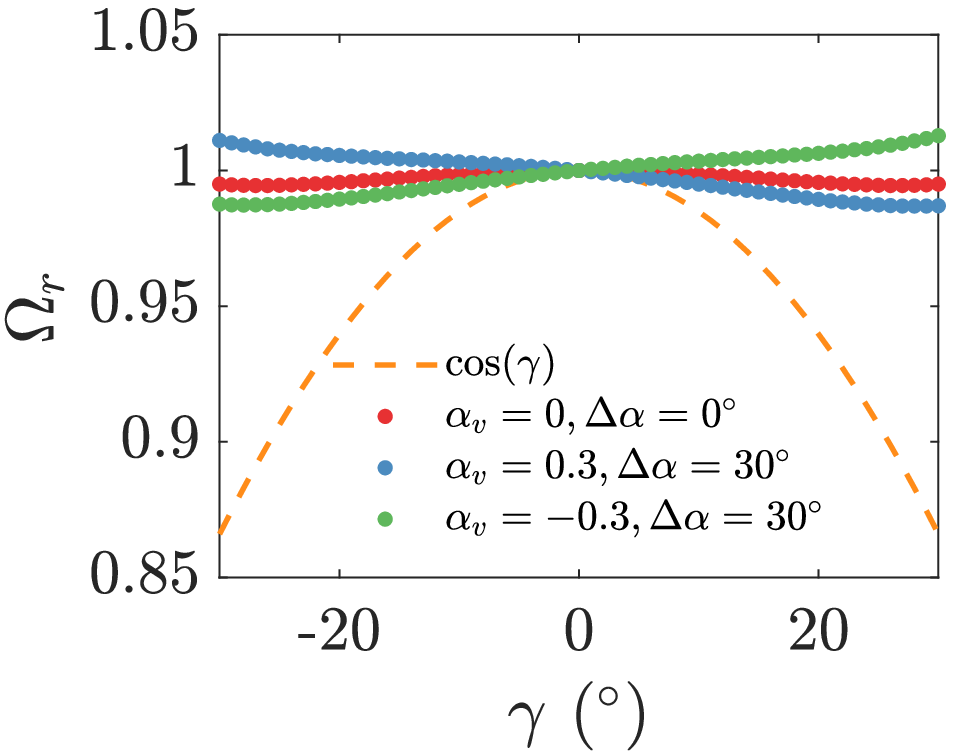} &
    \subfigimgtwo[width=\linewidth,valign=t]{(b)}{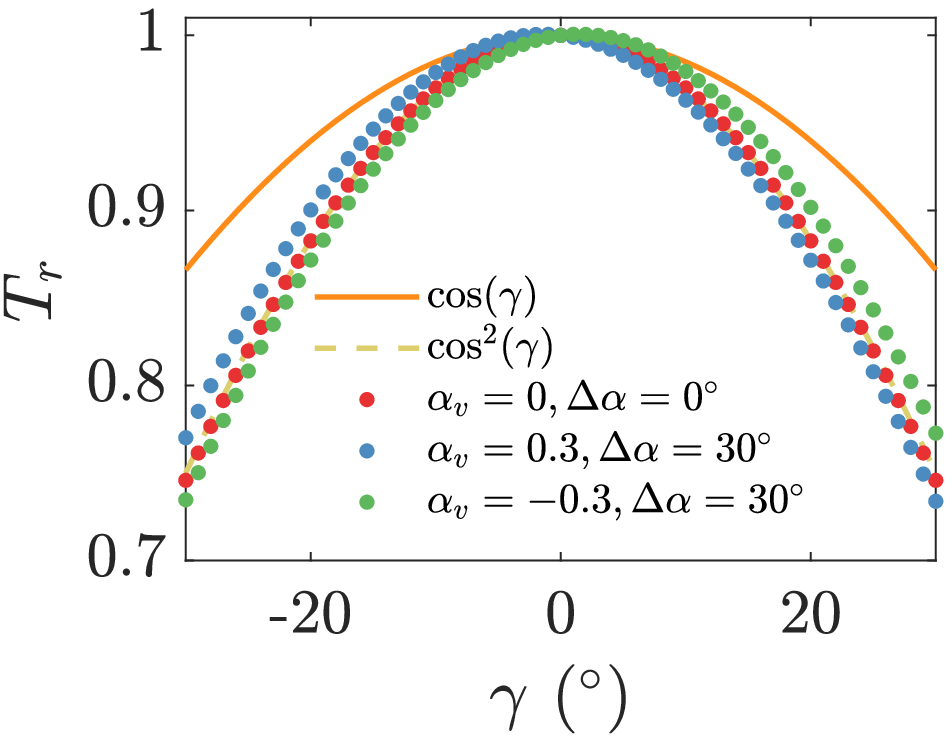} &
    \subfigimgtwo[width=\linewidth,valign=t]{(c)}{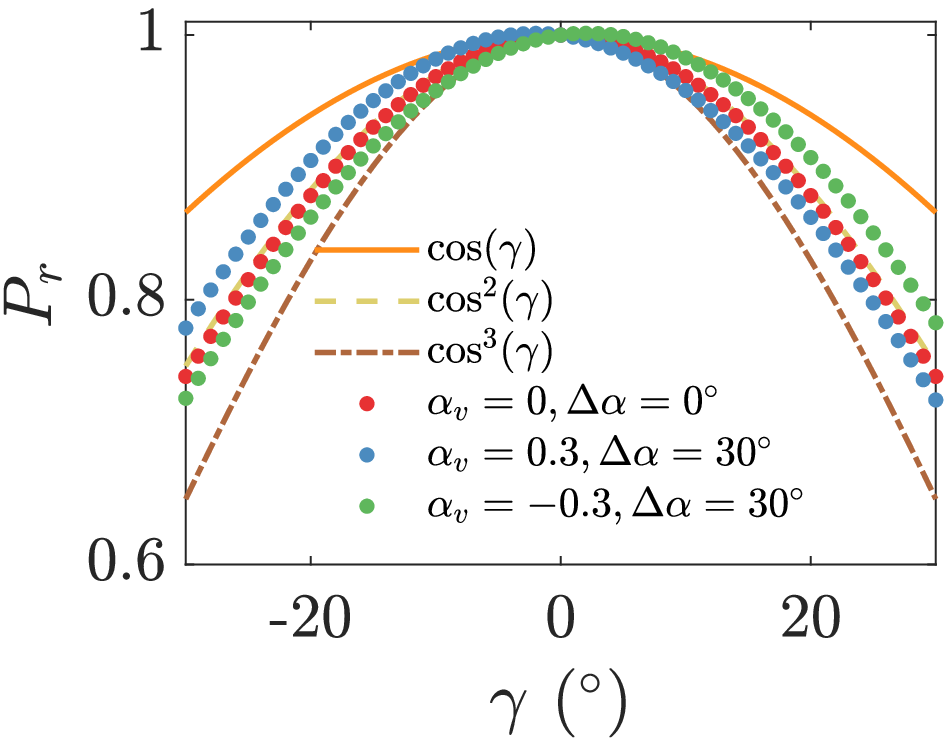} \\
    \subfigimgtwo[width=\linewidth,valign=t]{(d)}{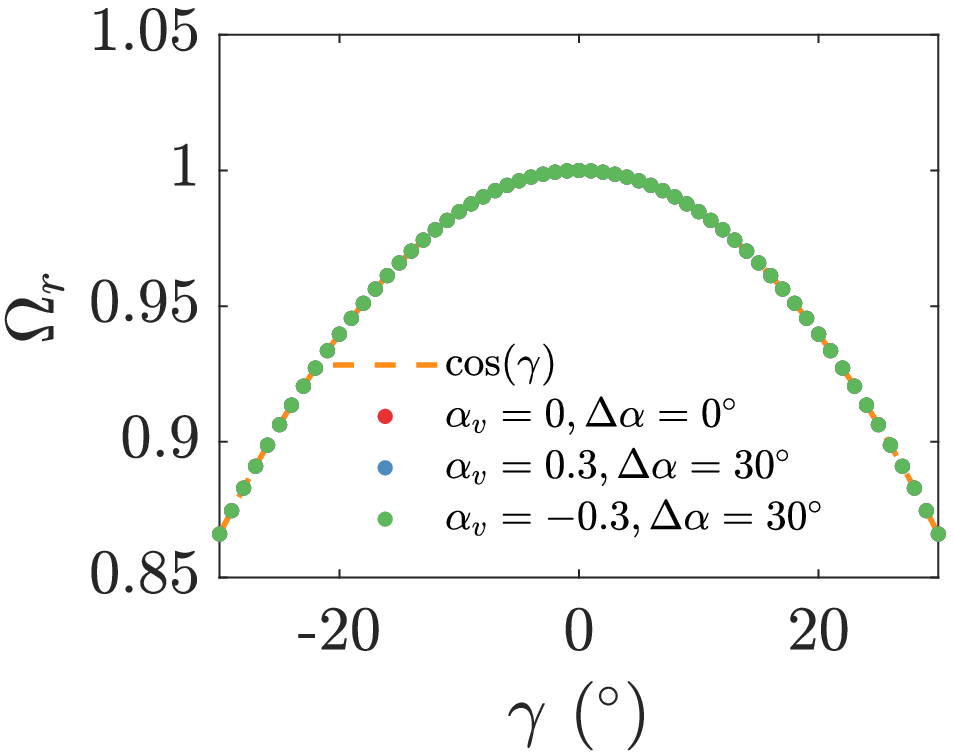} &
    \subfigimgtwo[width=\linewidth,valign=t]{(e)}{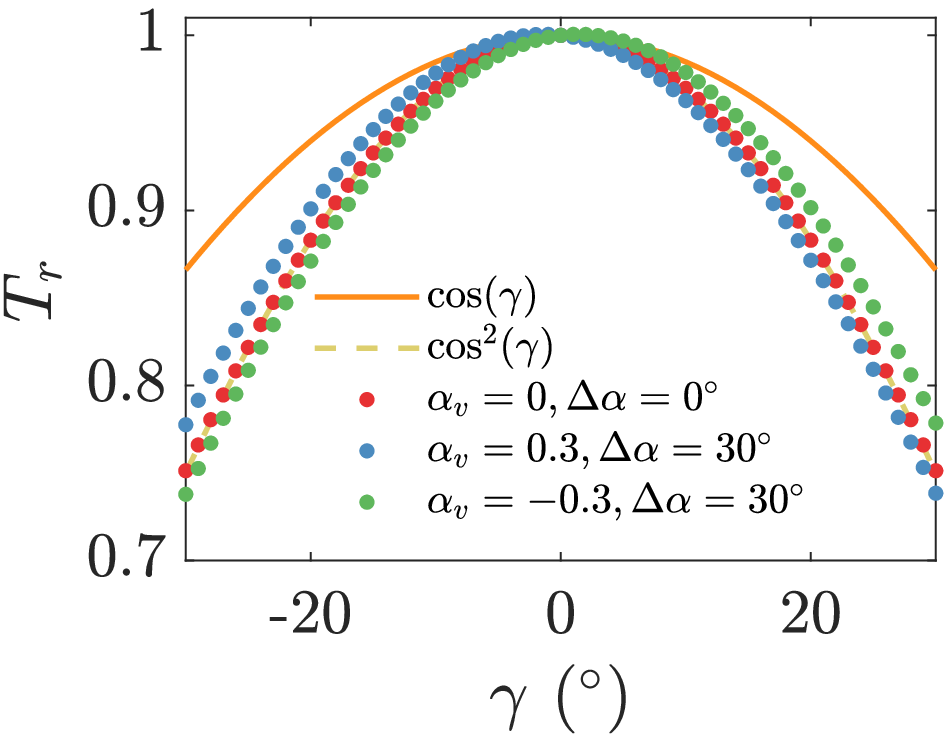} &
    \subfigimgtwo[width=\linewidth,valign=t]{(f)}{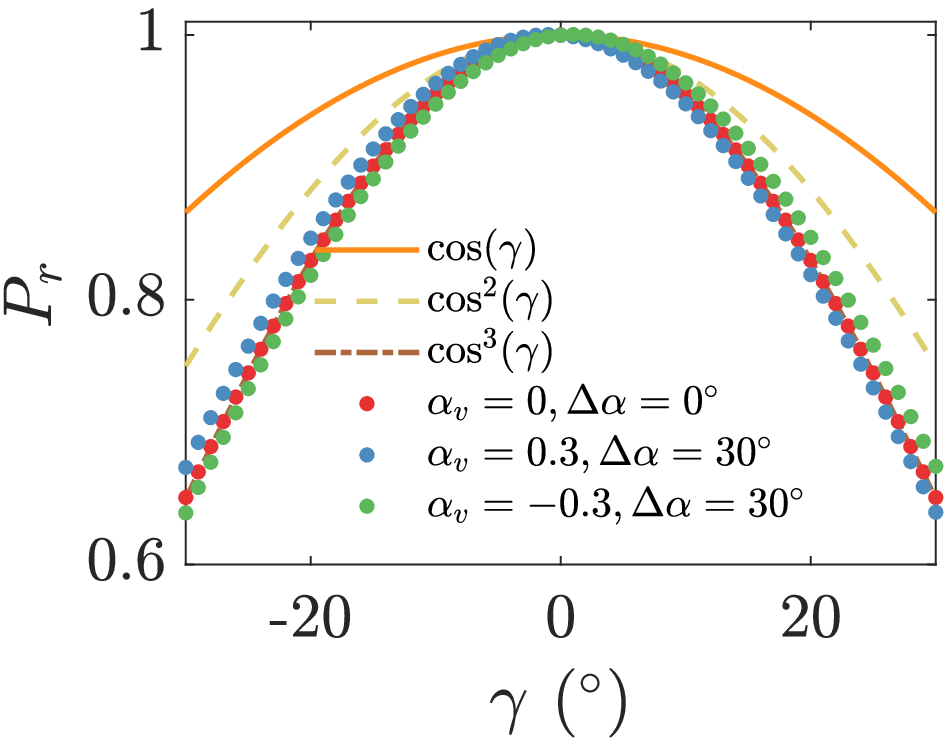} 
  \end{tabular}
  \caption{Model results with a power-law velocity profile and a linear veer profile.
  (a) $\Omega_r$, (b) $T_r$, and (c) $P_r$.
  The angular velocity $\Omega_\gamma$ is computed for each yaw misalignment angle by minimizing the difference between the aerodynamic and generator torques.
  (d-f) Same as (a-c) except the angular velocity ratio is prescribed as $\Omega_r = \Omega_\gamma/\Omega_{\gamma_0}= \cos(\gamma)$.}
    \label{fig:kragh_model_physics}
\end{figure}

The results from the model given by Eq. \ref{eq:pr} for a power law and linear veer profiles and the generator torque control described in \S \ref{sec:control} are shown in Figure \ref{fig:kragh_model_physics}(a-c).
For comparison, the power ratio model results for a prescribed $\Omega_r = \cos(\gamma)$ are shown for the same inflow wind profiles in Figure \ref{fig:kragh_model_physics}(d-f).
Considering the realistic control case with prescribed $K$, with no shear or veer $\alpha_v=\Delta \alpha=0,$ the resulting angular velocity ratio $\Omega_r\approx1$ for all yaw misalignment values. 
However, when shear and veer are incorporated, $\alpha_v=0.3, \Delta \alpha=30^\circ,$ asymmetry is introduced into $\Omega_r$, such that the angular velocity is higher for negative yaw misalignment than for positive yaw.
Correspondingly, the torque is also larger for negative yaw misalignment compared to positive yaw, and as a result, the power ratio $P_r$ is asymmetric, with $\gamma<0$ producing more power than $\gamma>0$.
This result agrees with the qualitative expectation that given positive shear and veer, there is more energy available above hub height than below hub height, and negative yaw misalignment will reduce the relative misalignment above the hub location.
Further, the power loss due to yaw misalignment cannot be approximated by a simple cosine model (as in Eq. \ref{eq:pr_general}), as the $P_r>\cos^2(\gamma)$ for $\gamma<0$ and $P_r<\cos^2(\gamma)$ for $\gamma>0$.
On the other hand, when $\alpha_v=-0.3$ and $\Delta \alpha=30^\circ$, the opposite qualitative asymmetry occurs, although the asymmetry is quantitatively different due to the asymmetric effects of the blade rotation direction.
While a power law form with $\alpha_v=-0.3$ is not likely to occur often in ABL observations, this result serves to approximate the influence of the sub-geostrophic jet which results in negative shearing conditions.

With $\Omega_r=\cos(\gamma)$, an asymmetry is present, but less pronounced.
Further, the power ratio approximately follows $\cos^3(\gamma)$ for all inflow profile cases.
This result confirms the expectation that the power ratio quantity $P_r(\gamma)$ will depend on the incident wind conditions and the control system specific to the wind turbine, although the model presented in \S \ref{sec:model_quant} can be used with arbitrary control laws or incident velocity profiles.
The model is compared to field experimental data in \S \ref{sec:results} using the control system for the presently studied wind turbines and the measured incident wind conditions.

\section{Wind farm and experimental setup}
\label{sec:exp}

\begin{figure}
  \centering
  \begin{tabular}{@{}p{0.45\linewidth}@{\quad}p{0.4\linewidth}@{}}
    \subfigimgtwo[width=\linewidth,valign=t]{(a)}{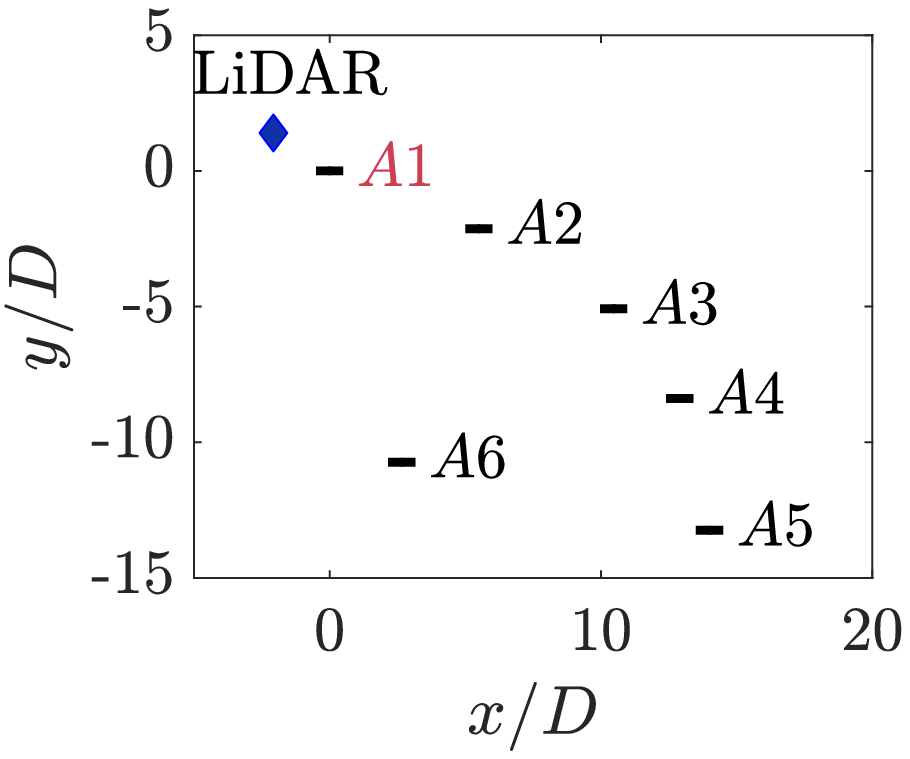} &
    \subfigimgfour[width=\linewidth,valign=t]{(b)}{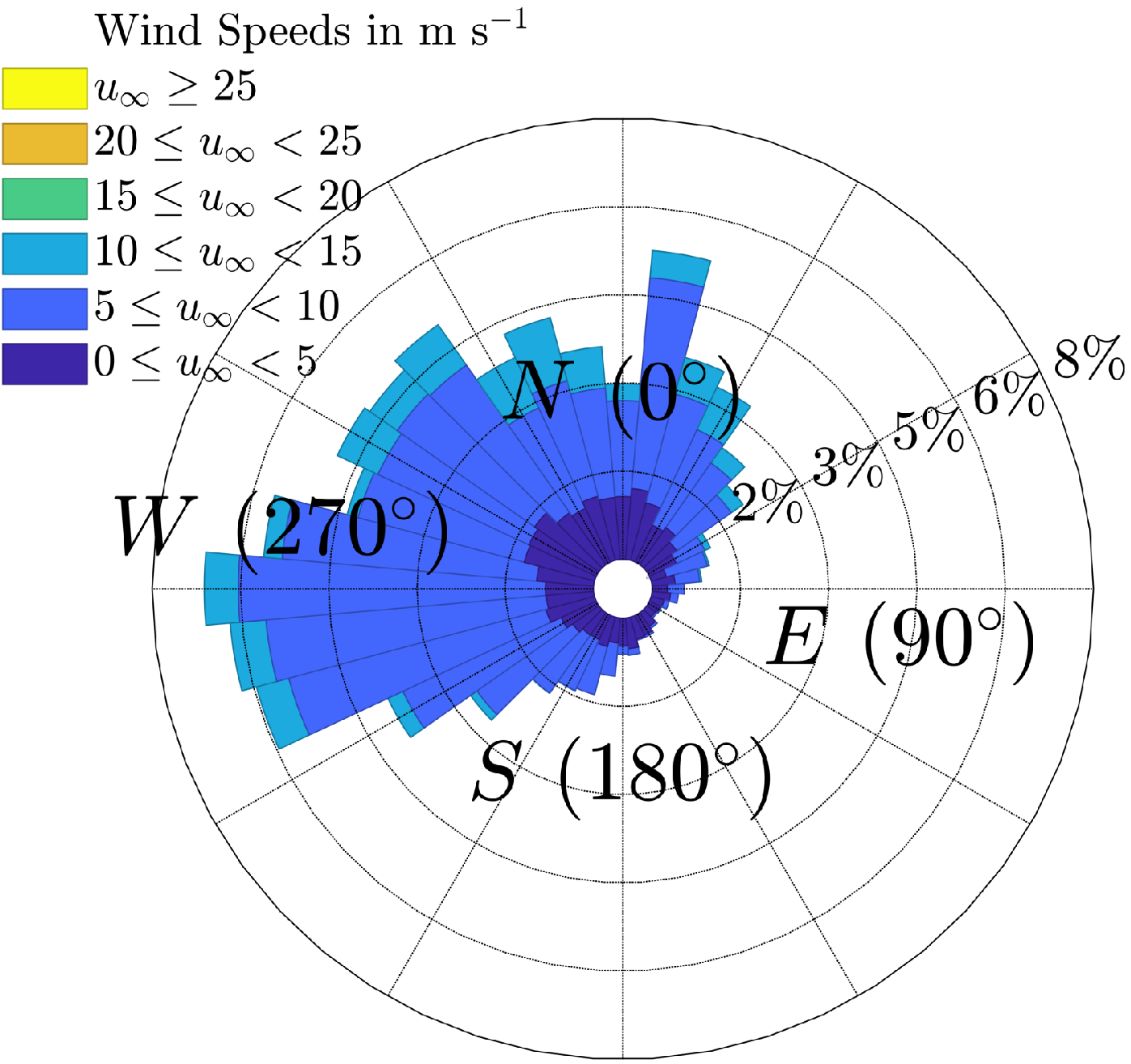}
  \end{tabular}
  \caption{(a) Wind turbine cluster A with the corresponding turbine labels.
  The easting and northing directions, normalized by the wind turbine diameter $D$, are indicted by $x$ and $y$, respectively.
  Red turbines are actuating turbines with intentional yaw misalignment strategies and black turbines are yaw aligned.
  The profiling LiDAR location, approximately $2.5D$ northwest of turbine $A1$, is indicated in cluster A.
  The wind turbines are oriented as if the flow is from the north.
  (b) Wind rose from the cluster A mast for February and March of 2016 through 2018.
        }
    \label{fig:topography}
\end{figure}

The wind farm of interest is located in northwest India.
The site contains nearly 100 utility-scale wind turbines of various original equipment manufacturer (OEM) construction.
The wind turbines have diameters and hub-heights of approximately 100 meters.
The wind farm topography is flat with a gradual slope increasing in elevation from northwest to southeast by approximately 100 meters over approximately 25 kilometers.
The wind turbine layout for the cluster of interest is shown in Figure \ref{fig:topography}(a).

The wind farm is characterized by two distinct wind seasons.
The summer corresponds to the Indian monsoon season \cite{gadgil2003indian} for which the wind speeds, as characterized by a site meteorological (MET) mast, are greater than $5$ and $10$ m/s approximately $87\%$ and $18\%$ of the time, respectively.
The prevalent wind direction during the summer wind season is from the southwest.
The other seasons are characterized by incident winds from the west, north, and northeast.
By contrast, the non-monsoon wind speeds are greater than $5$ and $10$ m/s approximately $70\%$ and $13\%$ of the time, respectively.

Cluster A (see Figure \ref{fig:topography}(a)) is in the northern region of the wind farm and is not affected by wind turbine wakes for flows that occur during the non-monsoon wind season.
The 2016-2018 yearly averaged monthly wind roses for the non-monsoon season measured by a MET mast approximately 20 kilometers west of cluster A are shown in Figure \ref{fig:topography}(b).

In order to measure the velocity profiles as a function of height incident on cluster A, a Leosphere Windcube V2.0 profiling LiDAR was installed at the field site.
The pulsed LiDAR measures backscatter by aerosols in the atmosphere and translates the measurements into a corresponding Doppler shift in order to provide information about the wind speed and direction.
The wind speed and direction have measurement uncertainties of $0.1$ m/s and $2^\circ$, respectively.
The measurement precisions for the wind speed and direction are $0.005$ m/s and $0.005^\circ$.
During the non-monsoon wind season, the LiDAR measures the wind speed and direction profiles upwind of turbine $A1$.
The profiling LiDAR measures the velocity at 12 range gates as a function of height between 43 and 200 meters of elevation.
A range gate is set at 104 meters to measure the velocity near hub height.

\begin{figure}
  \centering
  \begin{tabular}{@{}p{0.35\linewidth}@{\quad}p{0.35\linewidth}@{}}
  \subfigimgtwo[width=\linewidth,valign=t]{(a)}{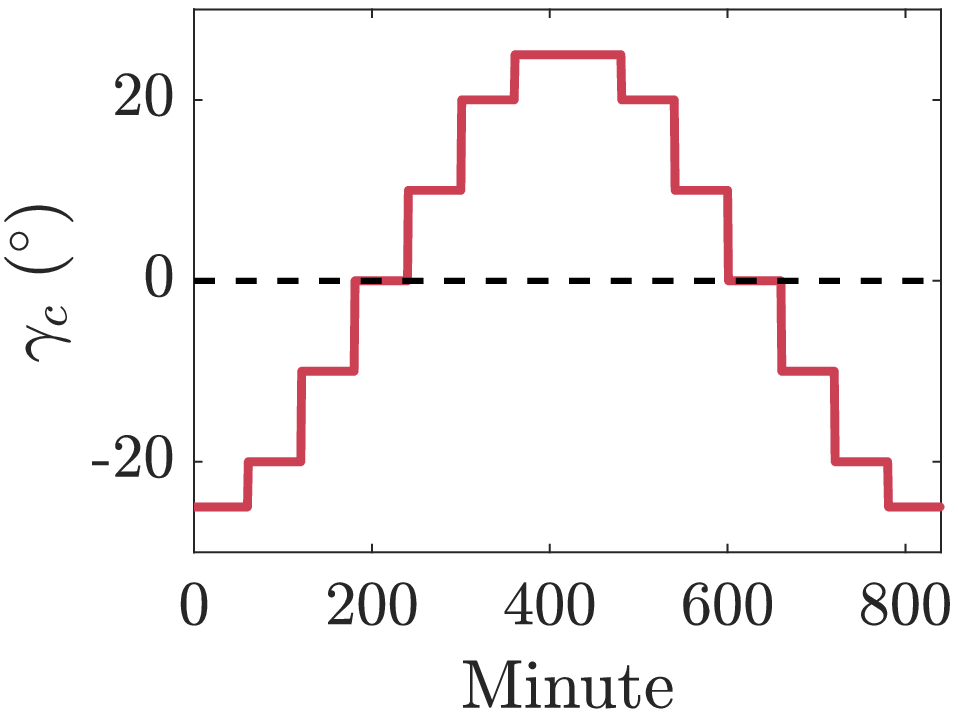} & 
  \subfigimgtwo[width=\linewidth,valign=t]{(b)}{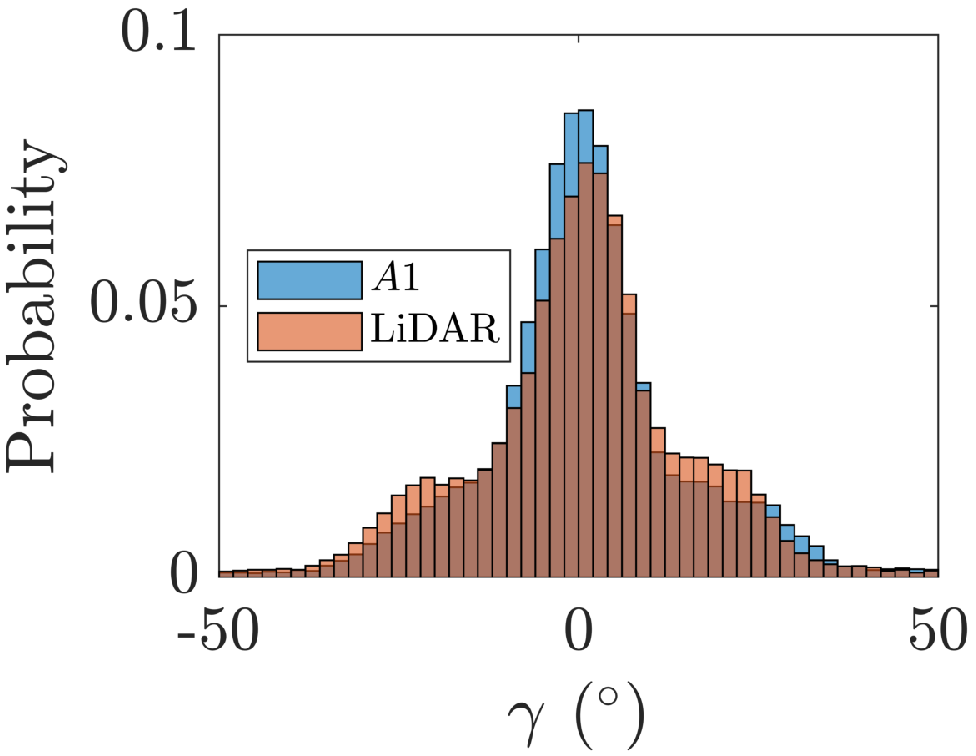}
  \end{tabular}
  \begin{tabular}{@{}p{0.70\linewidth}@{}}
    \subfigimgtwo[width=\linewidth,valign=t]{(c)}{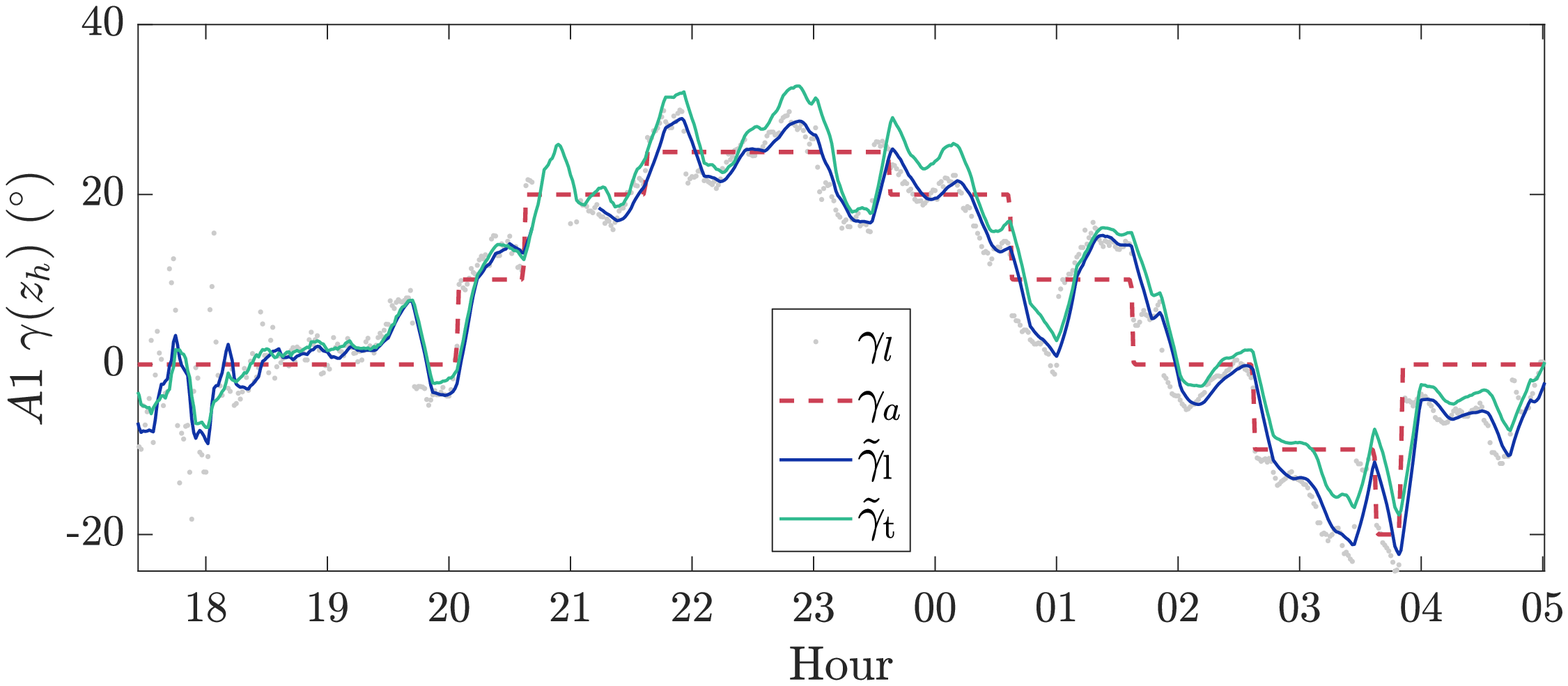}
  \end{tabular}
  \caption{
  (a) Time series of yaw offset $\gamma_c$ commands sent to each actuating wind turbine. 
  Each yaw misalignment command is held for one hour provided the wind conditions remain within the prescribed threshold parameters.
  (b) Probability distribution of the yaw misalignment calculated by turbine $A1$ and the profiling LiDAR.
  (c) Time series of turbine $A1$ yaw misalignment field experiment from March 19, 2020.
  The one-minute averaged yaw misalignment measured by the difference between the turbine nacelle position and the LiDAR wind direction is given by $\gamma_l = \alpha_{\mathrm{LiDAR}}(z=z_h) - \beta$, where $\beta$ is the nacelle position, the yaw misalignment applied by the turbine controller is $\gamma_a$, the ten-minute moving average of the yaw misalignment measured by the LiDAR is $\tilde{\gamma}_l$, and the ten-minute moving average of the yaw misalignment measured by turbine $A1$ is $\tilde{\gamma}_t$.
        }
    \label{fig:offset}
\end{figure}

The experiment was performed from February 12th, 2020 until April 7th, 2020.
In order to characterize the influence of yaw misalignment on freestream wind turbines, six full-scale wind turbines were provided with a yaw misalignment actuating sequence as a function of time (Figure \ref{fig:offset}(a)).
Turbine cluster A, the focus of this experiment given the nearby location of the LiDAR, with the actuating and reference turbines highlighted are shown in Figure \ref{fig:topography}(a).
For each turbine cluster, threshold wind condition parameters are set for which the yaw actuating time series would be followed.
The thresholds were prescribed as wind speeds in Region II of the turbine power curves as well as an arc of incident wind direction such that both the actuating and reference turbines are in freestream wind conditions with no upwind turbines within $20D$.
If the wind condition threshold values were violated, the actuating wind turbines SCADA commands $\gamma_c=0$, where $\gamma_c$ is the commanded yaw misalignment value.

The realized one-minute averaged yaw misalignment $\gamma_l = \alpha_{\mathrm{LiDAR}}(z=z_h) - \beta$ is computed as a difference between the wind turbine nacelle position and the LiDAR wind direction at hub height.
The yaw misalignment is also characterized by the wind turbine, where $\gamma_t$ is measured by a nacelle-mounted wind vane and reported as a relative wind direction with respect to the nacelle position orientation.
A histogram of the resulting yaw misalignment values for turbine $A1$, where the yaw misalignment is computed by the turbine ($\gamma_t$) or by the upwind profiling LiDAR ($\gamma_l$) is shown in Figure \ref{fig:offset}(b).
An example time series from the yaw misalignment field experiment is shown in Figure \ref{fig:offset}(c).
The SCADA applied yaw misalignment $\gamma_a$ attempts to follow Figure \ref{fig:offset}(a), provided the threshold conditions are met.
The LiDAR and wind turbine characterized ten-minute moving averaged yaw misalignment values, $\tilde{\gamma}_l$ and $\tilde{\gamma}_t$, where $\tilde{\cdot}$ denotes a ten-minute moving average, are also shown in Figure \ref{fig:offset}(c).
The wind vane relative wind direction measurement on the wind turbine nacelle is designed to measure the yaw misalignment offset for relatively low values of yaw.
The impact of yaw misalignment on the measurements of the nacelle mounted wind vane are uncertain, and therefore, in this study, the yaw misalignment will be characterized by the LiDAR wind direction and the wind turbine nacelle position $\gamma_l$.

\subsection{LiDAR measurements and stability quantification}
\label{sec:stability_quantification}

In order to establish a qualitative sense of the stability during the experiment, the bulk Richardson number is used \cite{stull2012introduction, zhan2020lidar}
\begin{equation}
Ri_B = \frac{(g/\overline{\theta}_T) \Delta \theta_T \Delta z}{(\Delta u)^2 + (\Delta v)^2}.
\end{equation}
The flow is statically unstable when $Ri_B<0$ and stable when $Ri_B>0$.
The magnitude of the bulk Richardson number indicates a qualitative sense of the dynamic stability of the flow, i.e. the balance between turbulent shear production and suppression of turbulence by stable stratification.
Critical values are not precisely defined as they are for the flux Richardson number, and therefore turbulence is expected even with $Ri_B \approx 10$ (see discussion in Stull (2012) \cite{stull2012introduction}).
The bulk Richardson number is computed with LiDAR measured velocity recorded at the wind turbine hub height, $z\approx100$ meters and at $z=43$ meters.
Temperature is measured at the ground by a LiDAR and at the 100-meter hub height by a nacelle-mounted thermometer.
Since the velocity and temperature measurements are not collocated and the $\Delta z$ layer is relatively thick compared to best practices \cite{stull2012introduction}, the bulk Richardson number computed in this study will only be used as a qualitative sense of stability.

\begin{figure}
  \centering
  \begin{tabular}{@{}p{0.33\linewidth}@{\quad}p{0.33\linewidth}@{\quad}p{0.33\linewidth}@{}}
    \subfigimgthree[width=\linewidth,valign=t]{(a)}{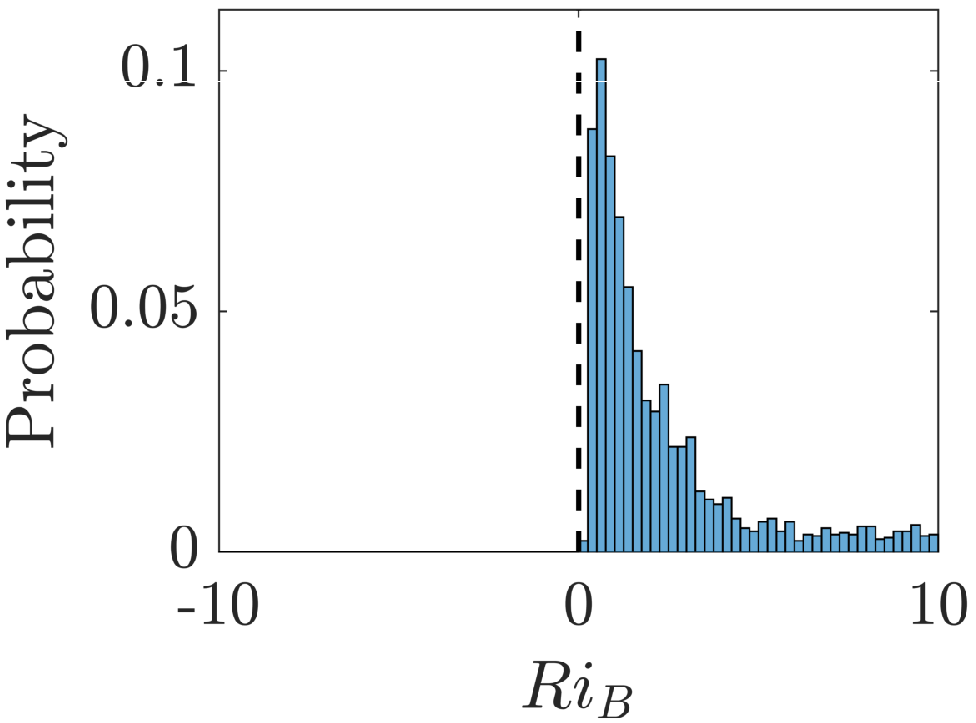} &
    \subfigimgthree[width=\linewidth,valign=t]{(b)}{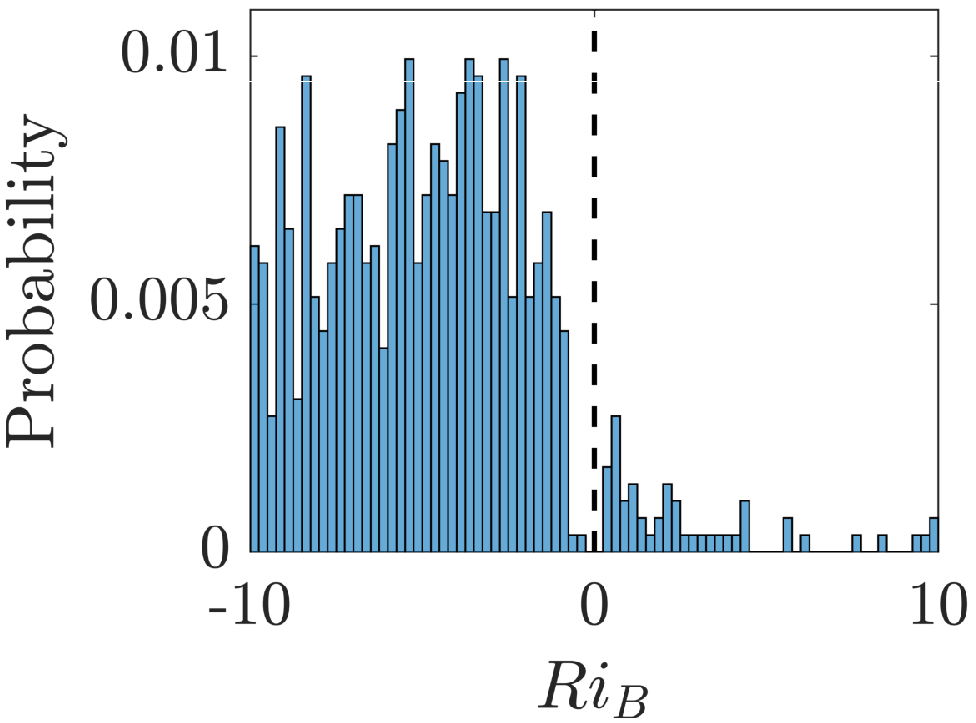} &
    \subfigimgthree[width=\linewidth,valign=t]{(c)}{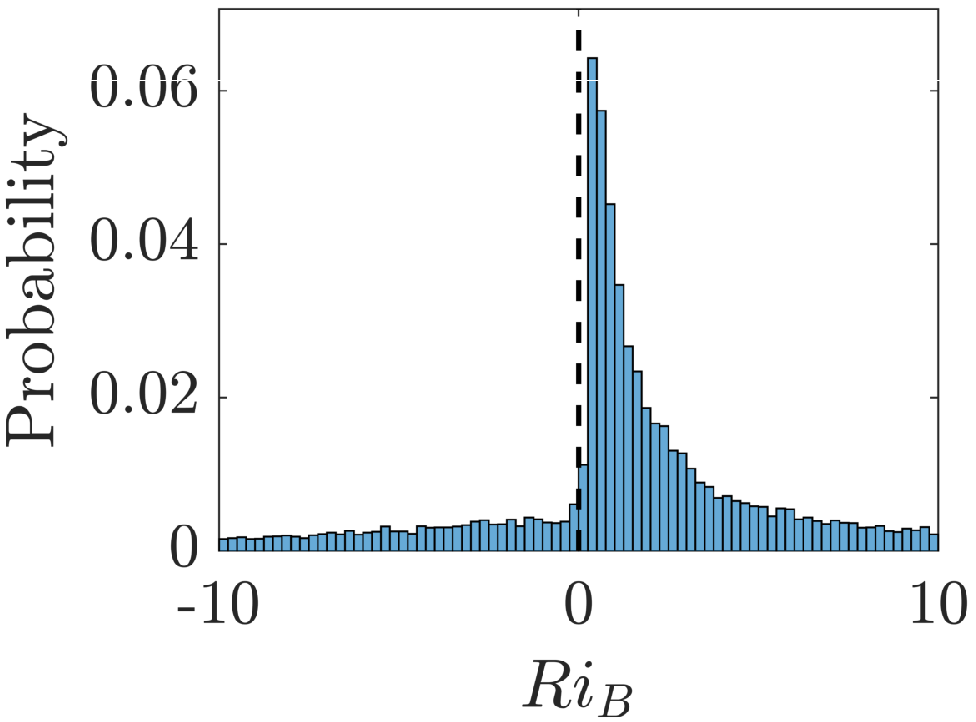}
  \end{tabular}
  \caption{Histograms of the bulk Richardson number $Ri_B$ for hour (a) 4 and (b) 14 and (c) for the flow of interest incident from $-30^\circ<\alpha<45^\circ$ for any hour.
  With $Ri_B<0$ the flow is statically unstable and with $Ri_B>0$, the flow is statically stable.
  $Ri_B = 0$ is shown with a vertical dashed black line.
  The same bin widths are used for each histogram.
        }
    \label{fig:stability_hist}
\end{figure}

The histograms of the bulk Richardson number for hours 4 and 14 of the day during the experiment are shown in Figure \ref{fig:stability_hist}(a,b).
In the early morning (hour 4), the ABL is statically stable with only positive bulk Richardson numbers.
Conversely, for hour 14, the ABL is generally unstable with $Ri_B<0$.
The flow of interest for the present experiment focuses on flow from the north between $-30^\circ < \alpha < 45^\circ$.
For flow constrained between these directions, the probability distribution of the bulk Richardson is shown in Figure \ref{fig:stability_hist}(c); the majority of the occasions of flow incident from the north results in statically stable ABL conditions ($Ri_B>0$).
Since most of the values of $Ri_B$ for the wind conditions of interest are positive and small, the flow will have shear turbulence production and be dynamically unstable but with a statically stable stratification which acts to suppress turbulent mixing.
As discussed in \S \ref{sec:stability}, the stable ABL with reduced turbulent mixing is expected to have stronger veer compared to the Ekman layer and a pronounced sub-geostrophic jet.

\begin{figure}
  \centering
  \begin{tabular}{@{}p{0.33\linewidth}@{\quad}p{0.33\linewidth}@{\quad}p{0.33\linewidth}@{}}
    \subfigimgtwo[width=\linewidth,valign=t]{(a)}{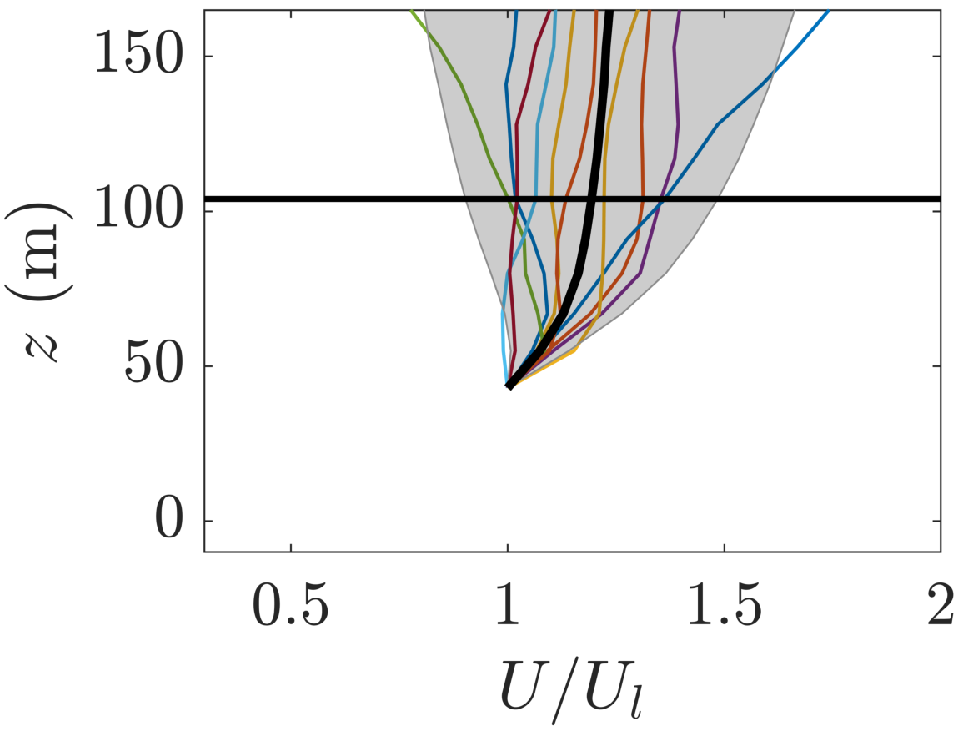} &
    \subfigimgtwo[width=\linewidth,valign=t]{(b)}{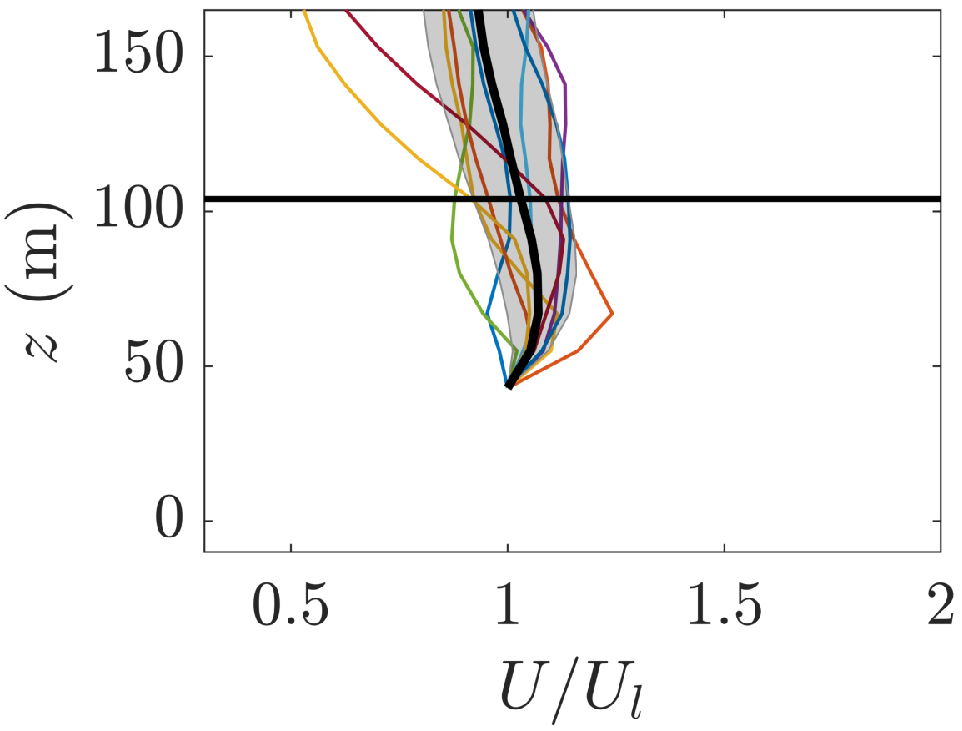} &
    \subfigimgtwo[width=\linewidth,valign=t]{(c)}{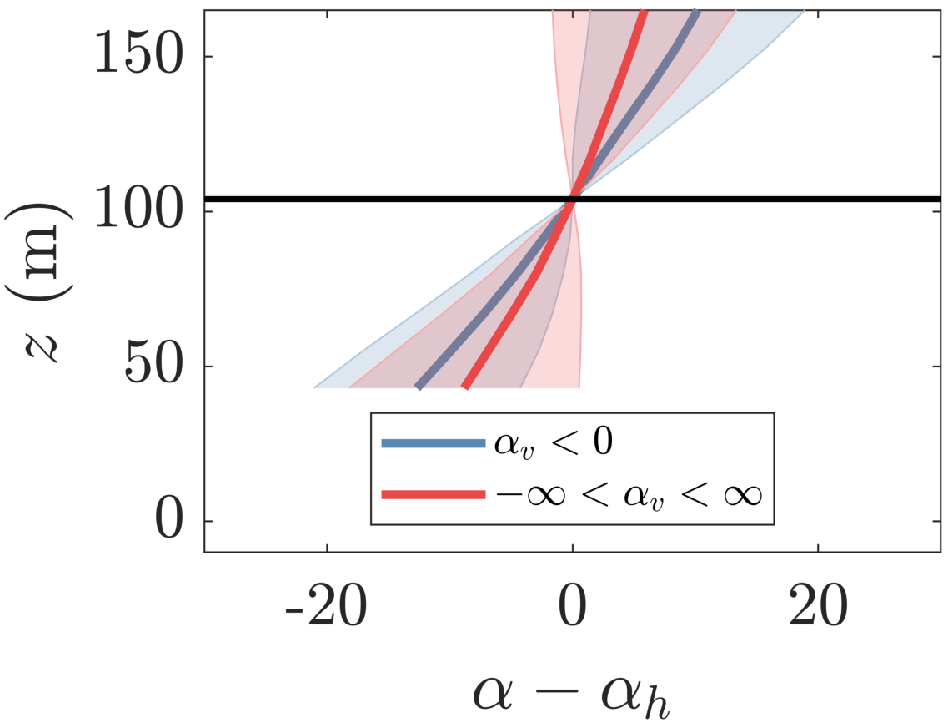}
  \end{tabular}
  \caption{Horizontal speed $U=\sqrt{u^2+v^2}$ normalized by the lowest wind speed measured by the profiling LiDAR $U_l$ at $z=43$ meters above the ground for (a) $-\infty<\alpha_v<\infty$ and (b) $\alpha_v<0$.
  The black curve represents the median with the shaded area representing one standard deviation about the median. Colored lines are 10 randomly selected one-minute averaged velocity profiles within the wind condition group.
  The horizontal black line indicates the wind turbine hub-height.
  (c) Median wind direction $\alpha-\alpha_h$ for the two directional shear cases in (a,b). The shaded error represents one standard deviation in the data.
        }
    \label{fig:shear}
\end{figure}

Given the expected stable ABL conditions during the experiment, the experimental results will be characterized in conditional averages depending on the magnitude of shear and veer recorded by the profiling LiDAR.
The LiDAR measures at 11 set-points vertically over the wind turbine rotor area.
The shear exponent $\alpha_v$ (Eq. \ref{eq:power_law}) is computed through a least-squares fit of the 11 points in the rotor area to a power law profile.
The median velocity profile for the northern flow experimental conditions for all values of $\alpha_v$ is shown in Figure \ref{fig:shear}(a).
Overlaid on the curve are 10 randomly selected one-minute averaged velocity profiles and the standard deviation about the median value as a function of $z$ is shown.
In the median, $\alpha_v=0.12$, but the standard deviation is substantial with negative shear occurring well within one standard deviation, indicating that the flow deviates from a power law with non-negligible frequency.
When the velocity profiles are filtered such that the least-squares error computed $\alpha_v<0$, a sub-geostrophic jet emerges in the median profile with some randomly selected profiles exhibiting significant anti-shear above the wind turbine hub-height (Figure \ref{fig:shear}(b)).
It is worth noting that a power law results in a poor fit to the velocity profiles in Figure \ref{fig:shear}(a,b), and therefore, $\alpha_v$ will be used only as a qualitative measure of the direction of shear in the wind profile. 
It is also evident from Figure \ref{fig:shear}(a,b) that the magnitude, and even the sign, of $\alpha_v$ is a function of $z$, which was also shown in a onshore wind farm in the Midwest of the United States \cite{sanchez2020effect}.

The median wind directions as functions of height for wind conditions filtered by $-\infty < \alpha_v < \infty$ and $\alpha_v<0$ are shown in Figure \ref{fig:shear}(c).
The median wind direction profiles are both increasing as a function of height, which is positive veering associated with Ekman turning (clockwise turning with increasing $z$).
The veer profiles are also approximately linear as a function of height, confirming the veer selections made in the canonical wind conditions model discussed in \S \ref{sec:model_quant} and shown in Figure \ref{fig:kragh_model_physics}.
When $\alpha_v<0$, the veer is significantly enhanced, with the median veer from the turbine bottom blade tip to top blade tip of $\Delta \alpha = \alpha(z=z_h+R)-\alpha(z=z_h-R)\approx 30^\circ$ compared to $\Delta \alpha \approx 15^\circ$ for the full $\alpha_v$ range.

\begin{figure}
  \centering
  \begin{tabular}{@{}p{0.33\linewidth}@{\quad}p{0.33\linewidth}@{\quad}p{0.33\linewidth}@{}}
    \subfigimgtwo[width=\linewidth,valign=t]{(a)}{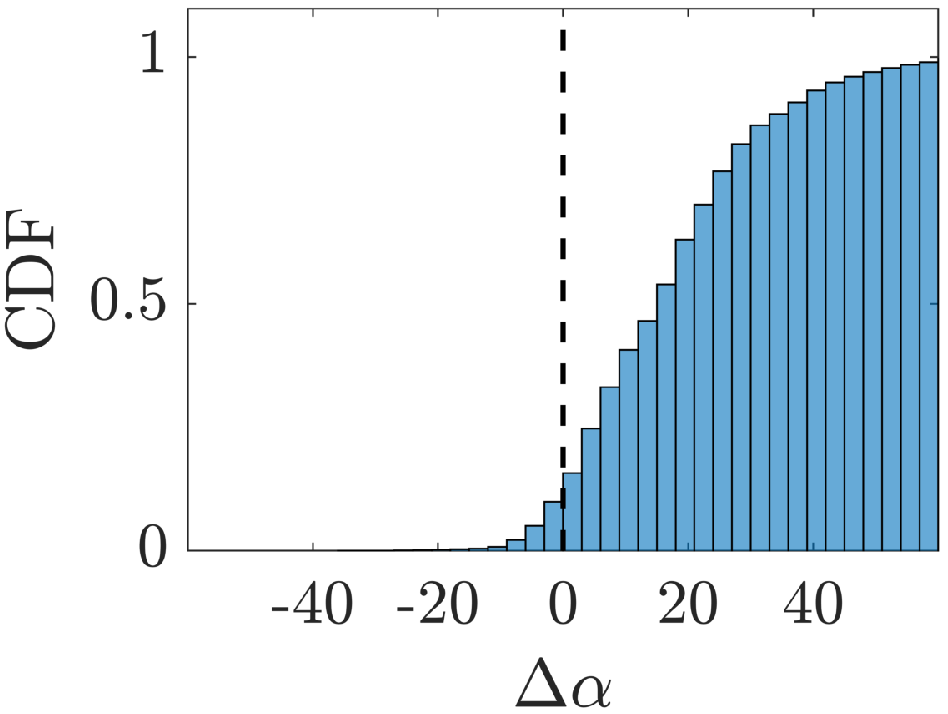} &
    \subfigimgtwo[width=\linewidth,valign=t]{(b)}{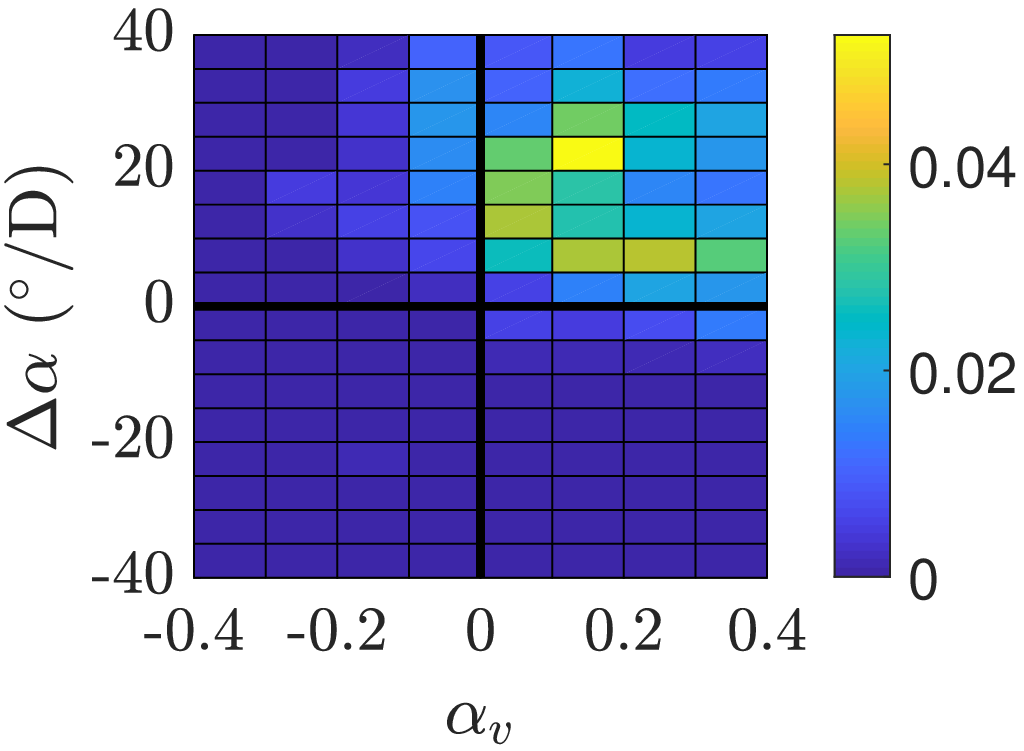} &
    \subfigimgtwo[width=\linewidth,valign=t]{(c)}{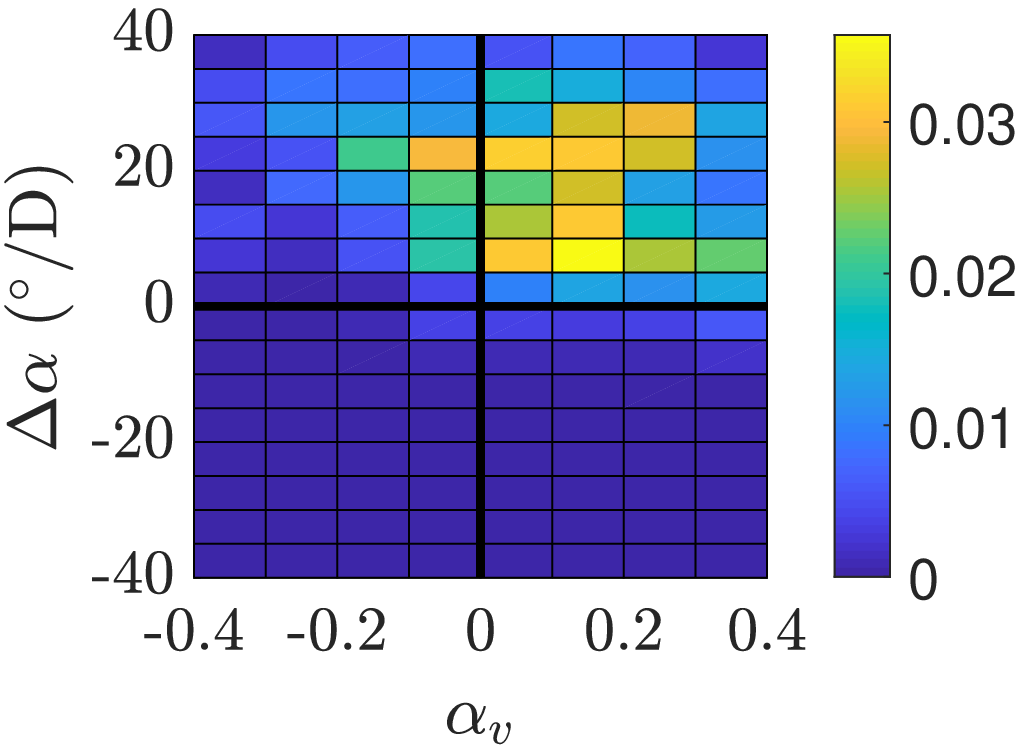}
  \end{tabular}
  \caption{(a) Probability distribution of the veer over the wind turbine face $\Delta \alpha$.
  (b) Joint probability distribution of the shear and veer measured for flow incident from the north.
  (c) Joint probability distribution of the shear above hub-height and veer measured for flow incident from the north.
        }
    \label{fig:veer_hist}
\end{figure}

The cumulative density function of the veer over the turbine face $\Delta \alpha$ for the wind conditions of interest is shown in Figure \ref{fig:veer_hist}(a).
Approximately $90\%$ of the one-minute averaged data samples have a positive veer and $10\%$ have negative veer (backing) which is similar to other field studies in flat terrain onshore wind farms (e.g. Sanchez \& Lundquist (2020) \cite{sanchez2020effect}).
Further, approximately $50\%$ of the veer cases result in $\Delta \alpha > 20^\circ$.
The joint probability distribution of $\Delta \alpha$ and $\alpha_v$ is shown in Figure \ref{fig:veer_hist}(b), for $\alpha_v$ computed using velocity measurements recorded between $43$ and $165$ m above the ground.
The majority of the one-minute averaged instances occur in quadrant 1, with $\alpha_v, \Delta \alpha>0$, and the following most frequent is quadrant 2, with $\alpha_v<0$ and $\Delta \alpha >0$.
As also shown in the cumulative distribution function in Figure \ref{fig:veer_hist}(a), $\Delta \alpha<0$ occurs infrequently.
The shear exponent is also computed considering vertical locations such that $z>z_h$ and the joint probability distribution is shown in Figure \ref{fig:veer_hist}(c).
Comparing Figures \ref{fig:veer_hist}(b) and (c), the frequency of $\alpha_v<0$ has significantly increased, indicating that the velocity profile above the wind turbine hub height often experiences negative shear with respect to the velocity at the wind turbine hub height.
Negative shear above hub height occurs approximately $35\%$ of the time and $\alpha_v<0.1$ occurs in $53\%$ of the one-minute averaged samples.

The power available in the incoming wind $P \propto (\vec{u} \cdot \hat{n})^3,$ where $\vec{u}$ is the incoming wind vector and $\hat{n}$ is the unit vector normal to the wind turbine rotor area.
The power available in the incoming wind will therefore depend on the specific wind speed and direction profiles.
Qualitatively, the probability distributions of $\alpha_v$ indicate that the available power in the wind is larger below hub height than above hub height with reasonable frequency at the wind farm in northwest India.

\section{Results}
\label{sec:results}

The power ratio ($P_r=P_{A1}/P_{A2}$, Equation \ref{eq:pr_general}), is calculated for the intentionally yaw misaligned turbine $A1$ with respect to the power production of baseline turbine $A2$ (see Figure \ref{fig:topography}(a) for the farm layout).
Turbine $A2$ is a turbine directly adjacent to $A1$ for wind incident from the north or northeast and the profiling LiDAR provides controlled wind measurements.
The influence of yaw misalignment on power production is computed as $P_r=P_{A1}/P_{A2}$ instead of $P_r=P_{A1}(\gamma)/P_{A1}(\gamma=0),$ to ensure that the same incident wind profile is encountered by the yaw misaligned and aligned turbines.

The wind conditions for cluster $A$ are restricted such that the wind direction at the wind turbine hub height is $-30^\circ<\alpha<45^\circ$ to ensure there is no waked inflow from potential upwind turbines outside the wind direction band of interest, as discussed in \S \ref{sec:exp}.
Additional quality filters are in place in the SCADA data to ensure that the wind turbines are operating normally with no power limitations, such as grid curtailment, and the yaw control system is active.
The turbulence intensity is constrained between $0\%<TI<10\%$ to reduce the variability in the wind conditions incident to turbines $A1$ and $A2$, although the results are similar with this constraint relaxed.

Given the experimental window of almost two months and the wind condition and data quality filters, $8,376$ unique, one-minute averaged data samples were collected which amounts to nearly 6 full days of yaw misalignment actuation spread over the two month period.
This results in approximately $700$ unique data points within each yaw misalignment offset command (Figure \ref{fig:offset}(a)).
As shown in Figure \ref{fig:offset}(c), due to the underlying dynamics of the native yaw control system, there are some deviations between the intended yaw misalignment and the realized yaw misalignment, as computed by the difference between the LiDAR wind direction at hub height and the wind turbine nacelle position.
The experimental $P_r$ results will therefore be analyzed with respect to the realized one-minute averaged yaw misalignment value, $\gamma_l = \alpha_{\mathrm{LiDAR}}(z=z_h) - \beta$ (Equation \ref{eq:gamma}) rather than the SCADA applied yaw value.
The mean Taylor's hypothesis advection time between the LiDAR and the wind turbines of interest is less than one minute.
The advection time lag is not included in the following analysis but the results are similar with an advection lag incorporated. 
Given the form of $P_r$ and that the wind speeds are restricted to Region II of the power curve, the particular value of the incident wind speed does not significantly influence $P_r$.

\begin{figure}
  \centering
  \begin{tabular}{@{}p{0.33\linewidth}@{\quad}p{0.33\linewidth}@{\quad}p{0.33\linewidth}@{}}
    \subfigimgthree[width=\linewidth,valign=t]{(a)}{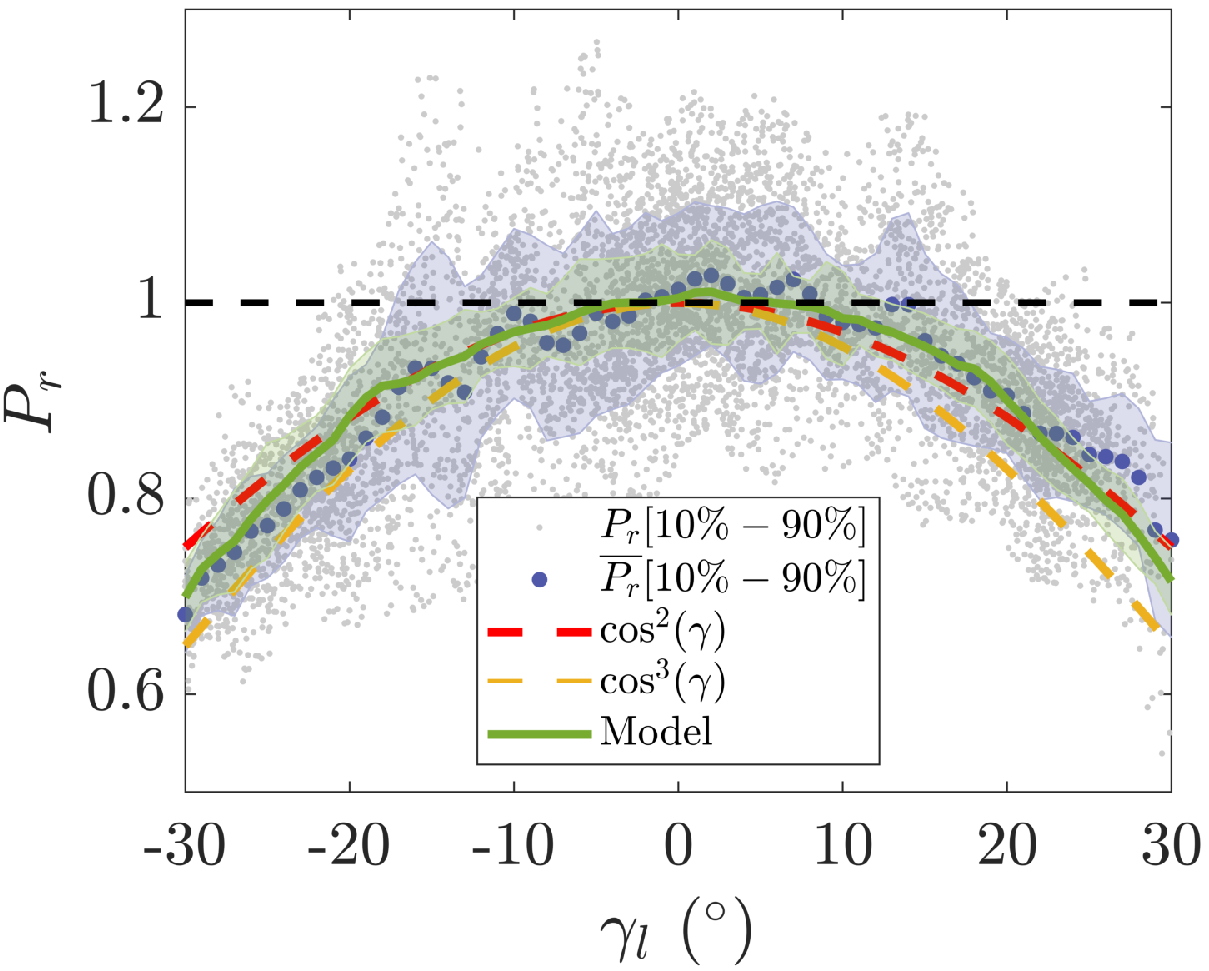} &
    \subfigimgthree[width=\linewidth,valign=t]{(b)}{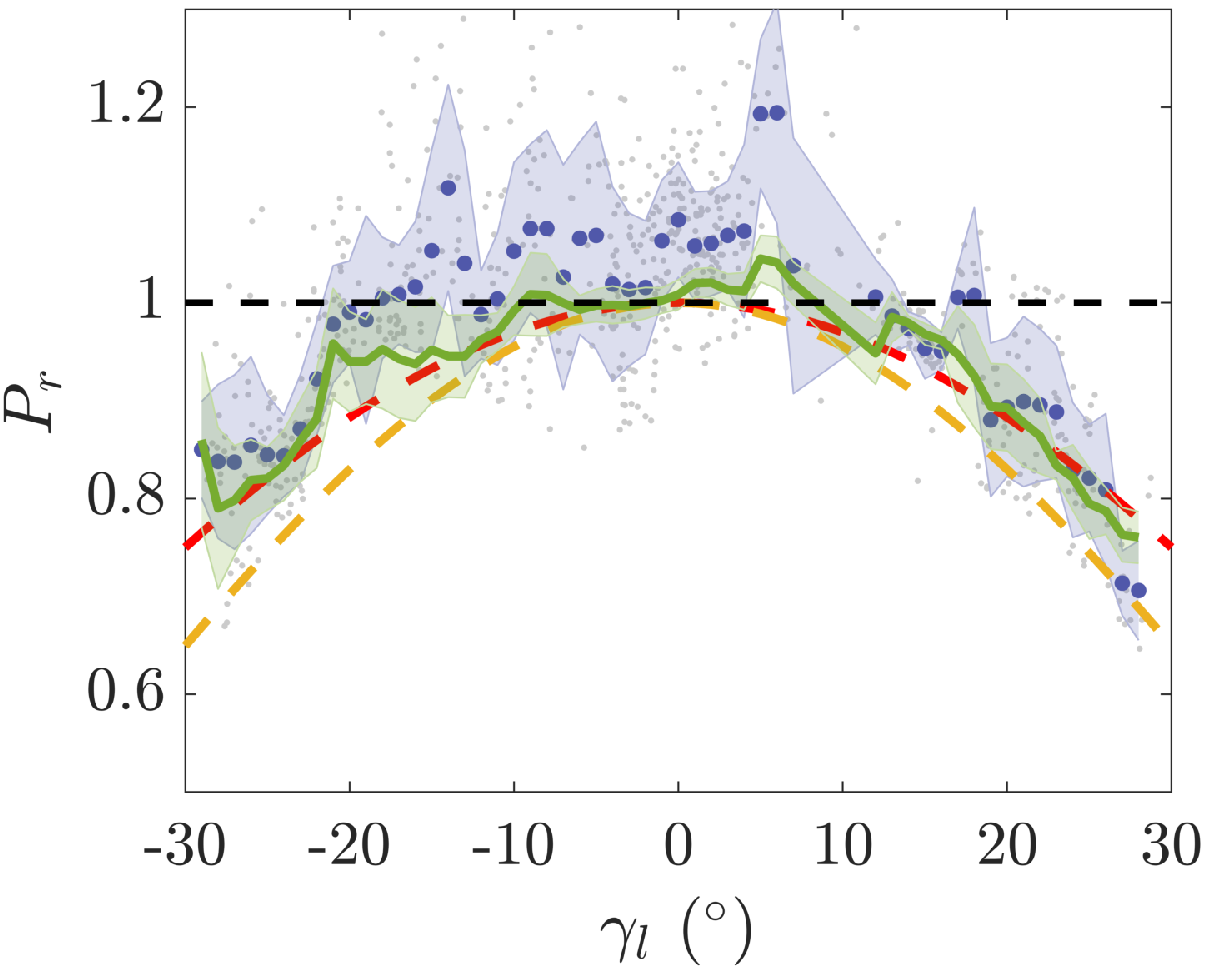} &
    \subfigimgthree[width=\linewidth,valign=t]{(c)}{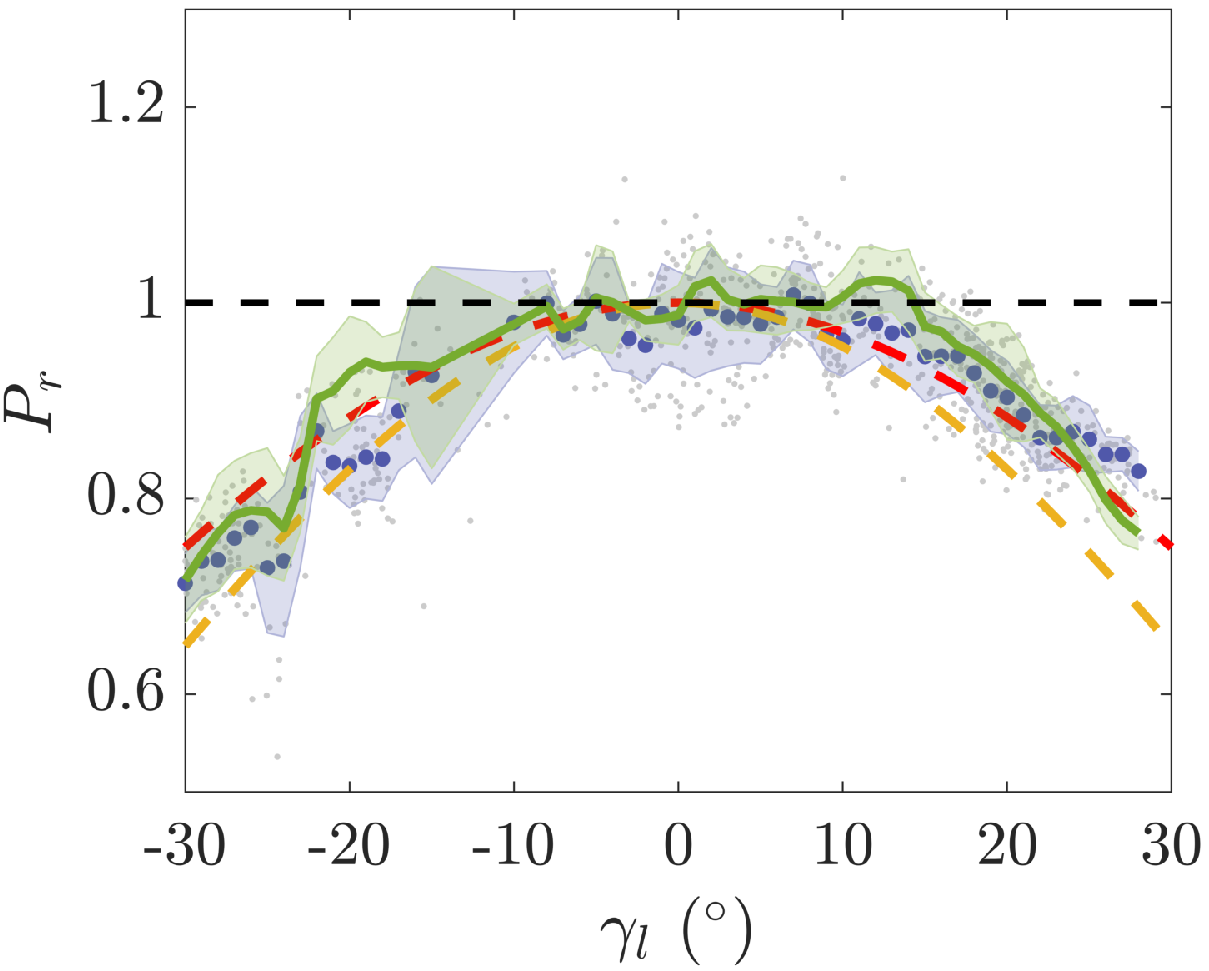}
    \end{tabular}
  \caption{Blade element model comparison. Power ratio $P_r = P_{A1}/P_{A2}$ for incident wind directions of $-30^\circ<\alpha(z=z_h)<45^\circ$ as measured by the LiDAR. 
  The yaw misalignment values are binned in $1^\circ$ increments.
  Within each yaw misalignment bin, $10\%$ tails on the upper and lower ends of the $P_r$ PDF are removed.
  $\overline{P_r}[10\%-90\%]$ denotes the median of the central $80\%$ of the $P_r$ PDF.
  The yaw misalignment $\gamma_l$ is calculated by the LiDAR $\gamma_l= \alpha_{\mathrm{LiDAR}}-\beta_{A1}$.
  (a) All conditions of shear and veer are considered and the turbulence intensity is constrained between $0<TI<10\%$. 
  The number of resulting data points is $n=8376$.
  (b) $\alpha_v>0.2$, $\Delta \alpha_v>20$, $n=873$
  (c) $\alpha_v<0$, $\Delta \alpha_v>20$, $n=996$.
  Conditional bins with more than $5$ data points are shown.
        }
    \label{fig:pr_full_data_physics}
\end{figure}

The power ratio $P_r$ for the full experimental dataset is shown in Figure \ref{fig:pr_full_data_physics}(a).
The realized yaw misalignment values $\gamma_l$ are binned in $1^\circ$ increments and the data within the middle $80\%$ of the probability distribution for each yaw misalignment bin are shown to alleviate the influence of one-minute averaged outliers.
The median of the middle $80\%$ is also shown with one standard deviation about the median representing the errorbars.
Reference curves for $\cos^2(\gamma)$ and $\cos^3(\gamma)$ are also shown.
Finally, the model presented in \S \ref{sec:model_quant} (Equation \ref{eq:pr}) is computed given the incident wind speed and direction profiles measured by the LiDAR for each one-minute average sample. 
The torque controller generator torque is prescribed as $T_c=K\Omega^2$.
The median and standard deviation about the median for the model are also shown in Figure \ref{fig:pr_full_data_physics}.

For the full experimental dataset (Figure \ref{fig:pr_full_data_physics}(a)), the power ratio approximately follows $\cos^2(\gamma)$ although $\cos^3(\gamma)$ is generally within one standard deviation of the data.
The $P_r$ is asymmetric, with $P_r(\gamma_l>0)>Pr(\gamma_l<0)$ for a fixed absolute value of $\gamma_l$.
With $\gamma_l>0$, $\cos^2(\gamma)$ is an underestimate of the $P_r$ but always remains within one standard deviation of the median value.
The curve for $\cos^3(\gamma)$ is also within one standard deviation of the median for $\gamma_l>0$ except for high values of $\gamma_l>25^\circ$.
On the contrary, for $\gamma_l<0$, $\cos^2(\gamma)$ is an overestimate of $P_r$ and falls outside of one standard deviation around the median for $\gamma_l<-25^\circ$.
For $\gamma_l<0$, $\cos^3(\gamma)$ is always within one standard deviation of the median.
These results reflect the expectation that $P_r$ will be asymmetric about $\gamma_l=0$ for spatially heterogeneous flow conditions in $z$.
The model generally follows $\cos^2(\gamma)$ with a slight deviation and asymmetry present; the model predicts that $\gamma_l>0$ produces slightly higher values of $P_r$ than $\gamma_l<0$ as the data also represents.
In order to account for potential causes of the asymmetry in $P_r$ as a function of $\gamma_l$, we will introduce wind condition restrictions on the full, recorded dataset.

In Figure \ref{fig:pr_full_data_physics}(b), $\alpha_v>0.2$ and $\Delta \alpha>20^\circ$ and there is a significant modification to the $P_r$ results.
In particular, there is a significant increase in $P_r(\gamma_l<0)$ and moderate reduction in $P_r(\gamma_l>0)$.
Within these conditions, $\cos^2(\gamma)$ is an underestimate of $P_r(\gamma_l<0)$, compared to the previous results considering all $\alpha_v$ where $\cos^2(\gamma)$ overestimated $P_r(\gamma_l<0)$.
With a positive veering angle associated with clockwise Ekman spiraling, a negative hub height yaw misalignment results in a smaller relative local yaw misalignment angle (Equation \ref{eq:gamma_local}) above hub height than below hub height.
With a strong positive shear exponent, $\alpha_v>0.2$, the wind speed also increases as a function of $z$.
Therefore, the local available power $(\vec{u}\cdot\hat{n})^3$ will be larger for a hub height yaw misalignment of $\gamma<0$ than for $\gamma>0$.
The model proposed in this study is able to capture the qualitative trend observed in the data where $P_r(\gamma_l<0)>P_r(\gamma_l>0)$.

In Figure \ref{fig:pr_full_data_physics}(c), the wind conditions are restricted to $\Delta \alpha>20^\circ$ and $\alpha_v<0$.
Given these wind conditions, there is an increase in $P_r(\gamma_l>0)$ and a reduction in $P_r(\gamma_l<0)$.
For negative shearing conditions, there is, in general, more energy below the wind turbine hub height of $z=z_h$ than above it.
Again, given $\Delta \alpha>20^\circ$, a positive yaw misalignment angle will locally align the rotor area with the inflow below hub height, and therefore, $P_r(\gamma_l>0)>P_r(\gamma_l<0)$ is expected.
Further, since the veering angle is more significant in negative shearing conditions (as discussed in \S \ref{sec:stability} and shown in Figure \ref{fig:shear}(c)), the reduction in $P_r$ for $\gamma_l<0$ is expected to be more substantial than the reduction in $P_r$ for $\gamma_l>0$ when $\alpha_v>0$.
In Figure \ref{fig:pr_full_data_physics}(c), there are sharp reductions in the $P_r$ for certain instances of $\gamma_l<0,$ confirming this expectation.
Again, the model captures the qualitative trend in $P_r$ although some quantitative discrepancies exist.

\begin{figure}
  \centering
  \begin{tabular}{@{}p{0.33\linewidth}@{\quad}p{0.33\linewidth}@{\quad}p{0.33\linewidth}@{}}
    \subfigimgthree[width=\linewidth,valign=t]{(a)}{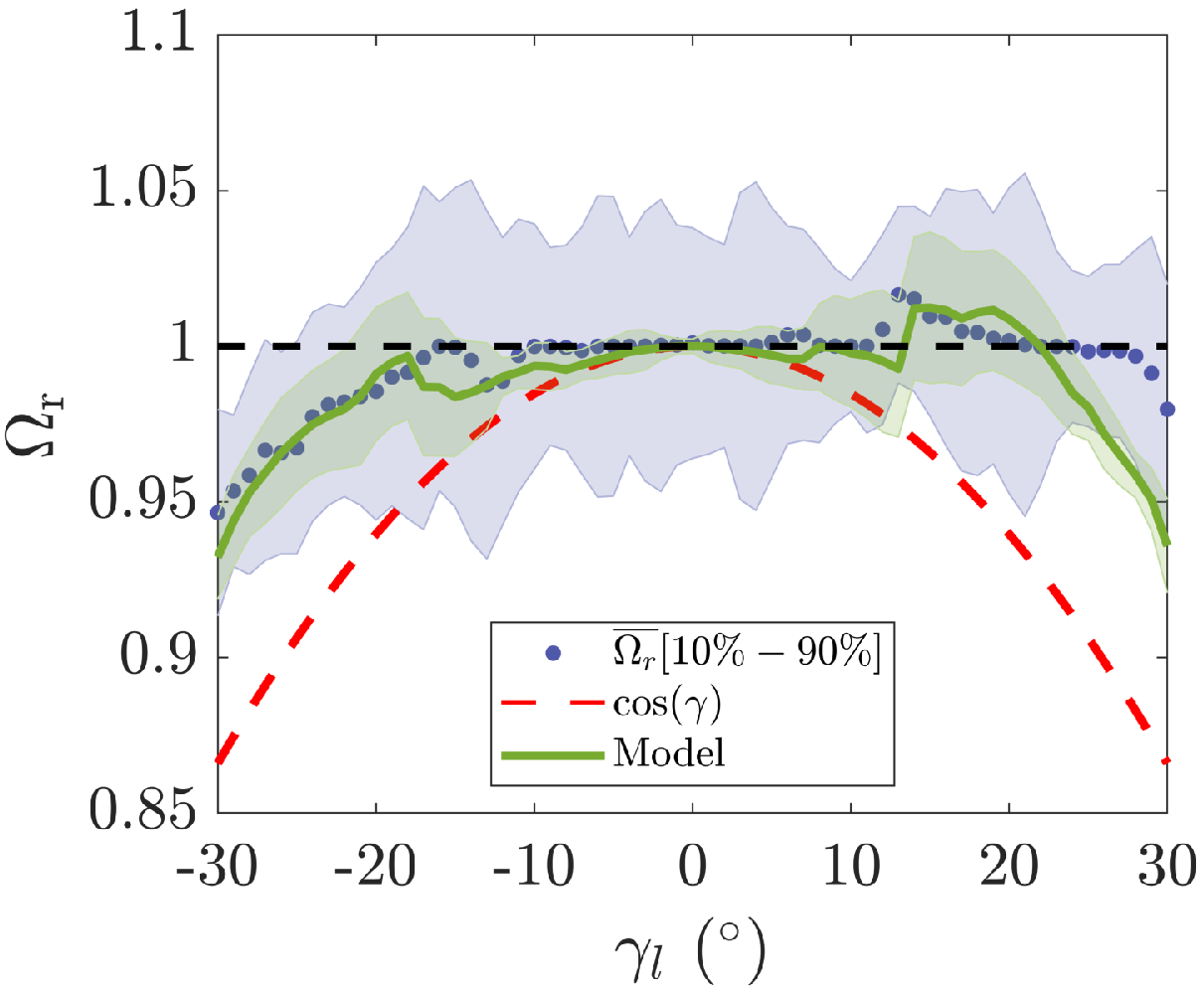} &
    \subfigimgthree[width=\linewidth,valign=t]{(b)}{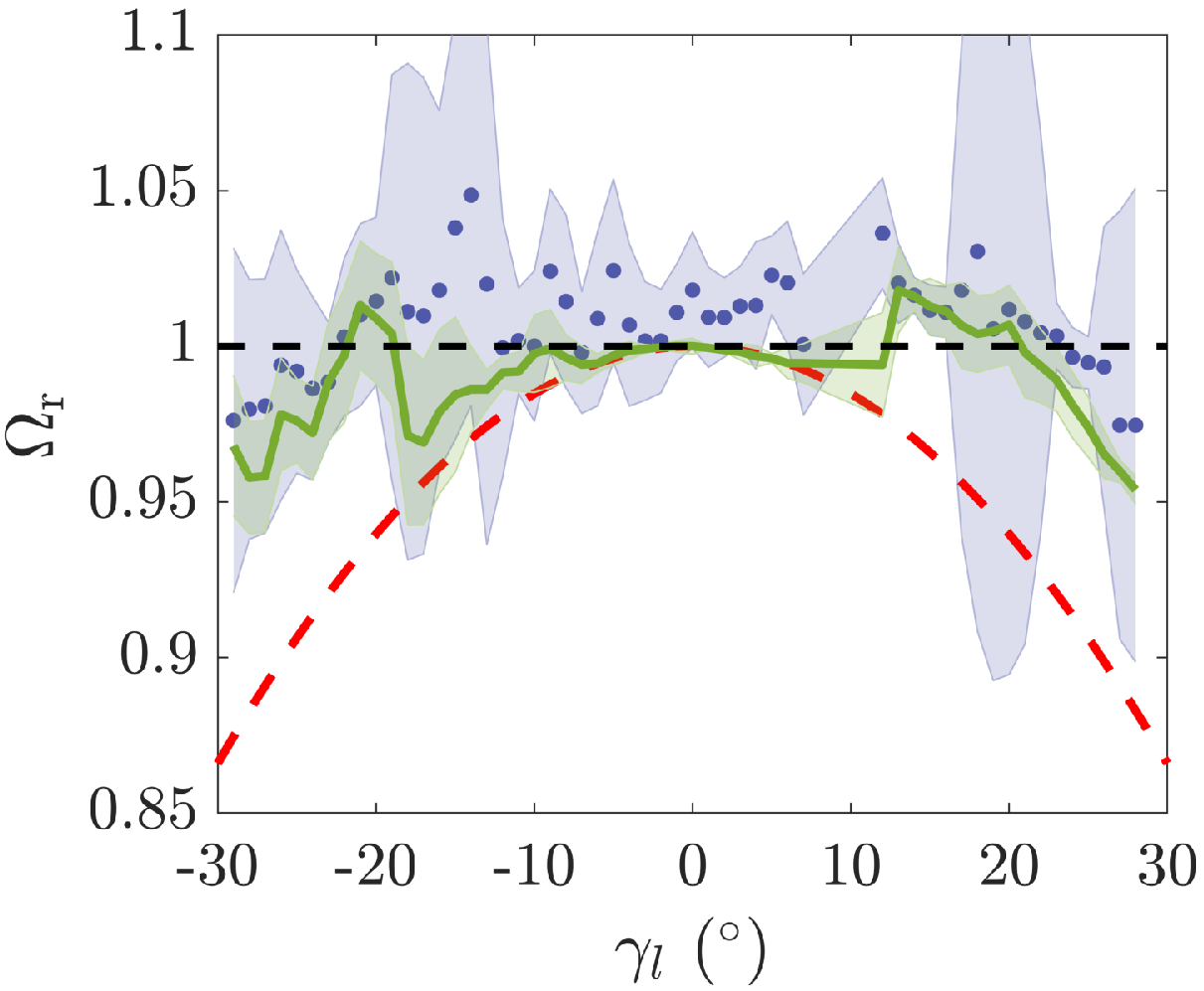} &
    \subfigimgthree[width=\linewidth,valign=t]{(c)}{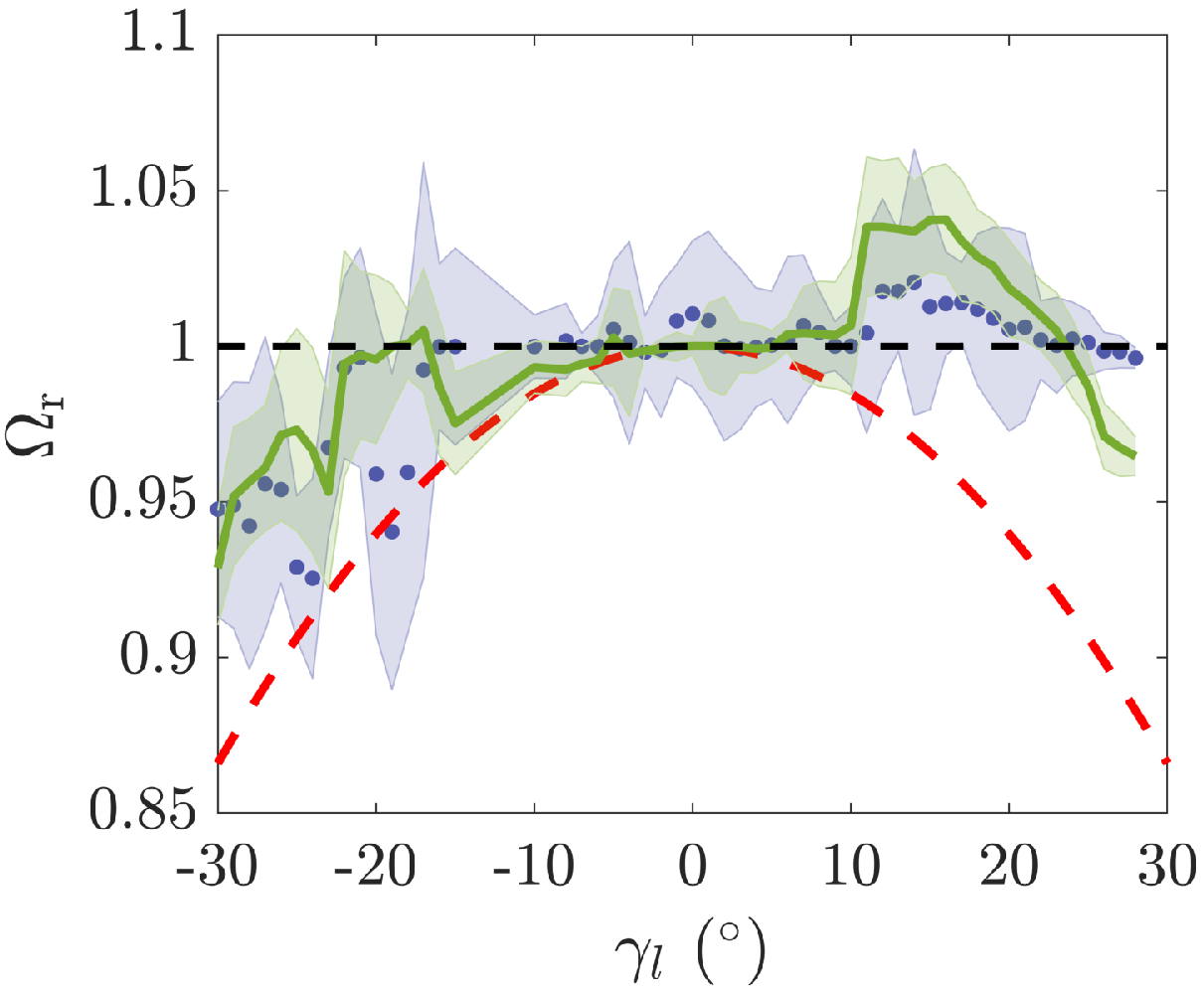}
    \end{tabular}
  \caption{Blade element model comparison. Angular velocity ratio $\Omega_r = \Omega_{A1}/\Omega_{A2}$ for incident wind directions of $-30^\circ<\alpha(z=z_h)<45^\circ$ as measured by the LiDAR. 
  The yaw misalignment values are binned in $1^\circ$ increments.
  Within each yaw misalignment bin, $10\%$ tails on the upper and lower ends of the $\Omega_r$ PDF are removed.
  $\overline{\Omega_r}[10\%-90\%]$ denotes the median of the central $80\%$ of the $\Omega_r$ PDF.
  The yaw misalignment $\gamma_l$ is calculated by the LiDAR $\gamma_l= \alpha_{\mathrm{LiDAR}}-\beta_{A1}$.
  (a) All conditions of shear and veer are considered and the turbulence intensity is constrained between $0<TI<10\%$. 
  The number of resulting data points is $n=8376$.
  (b) $\alpha_v>0.2$, $\Delta \alpha_v>20$, $n=873$
  (c) $\alpha_v<0$, $\Delta \alpha_v>20$, $n=996$.
  Conditional bins with more than $5$ data points are shown.
        }
    \label{fig:omega}
\end{figure}

\begin{figure}
  \centering
  \begin{tabular}{@{}p{0.33\linewidth}@{\quad}p{0.33\linewidth}@{\quad}p{0.33\linewidth}@{}}
    \subfigimgthree[width=\linewidth,valign=t]{(a)}{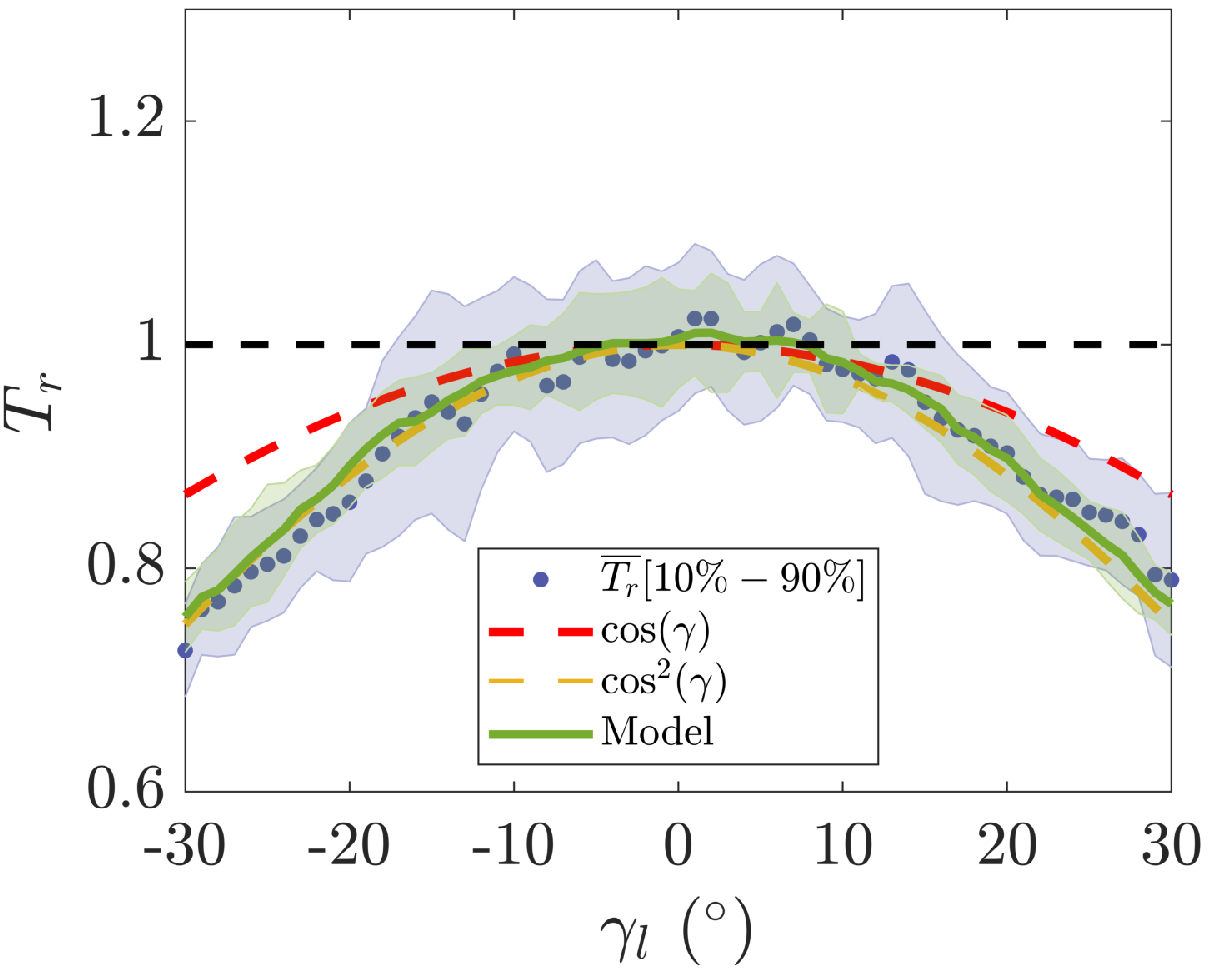} &
    \subfigimgthree[width=\linewidth,valign=t]{(b)}{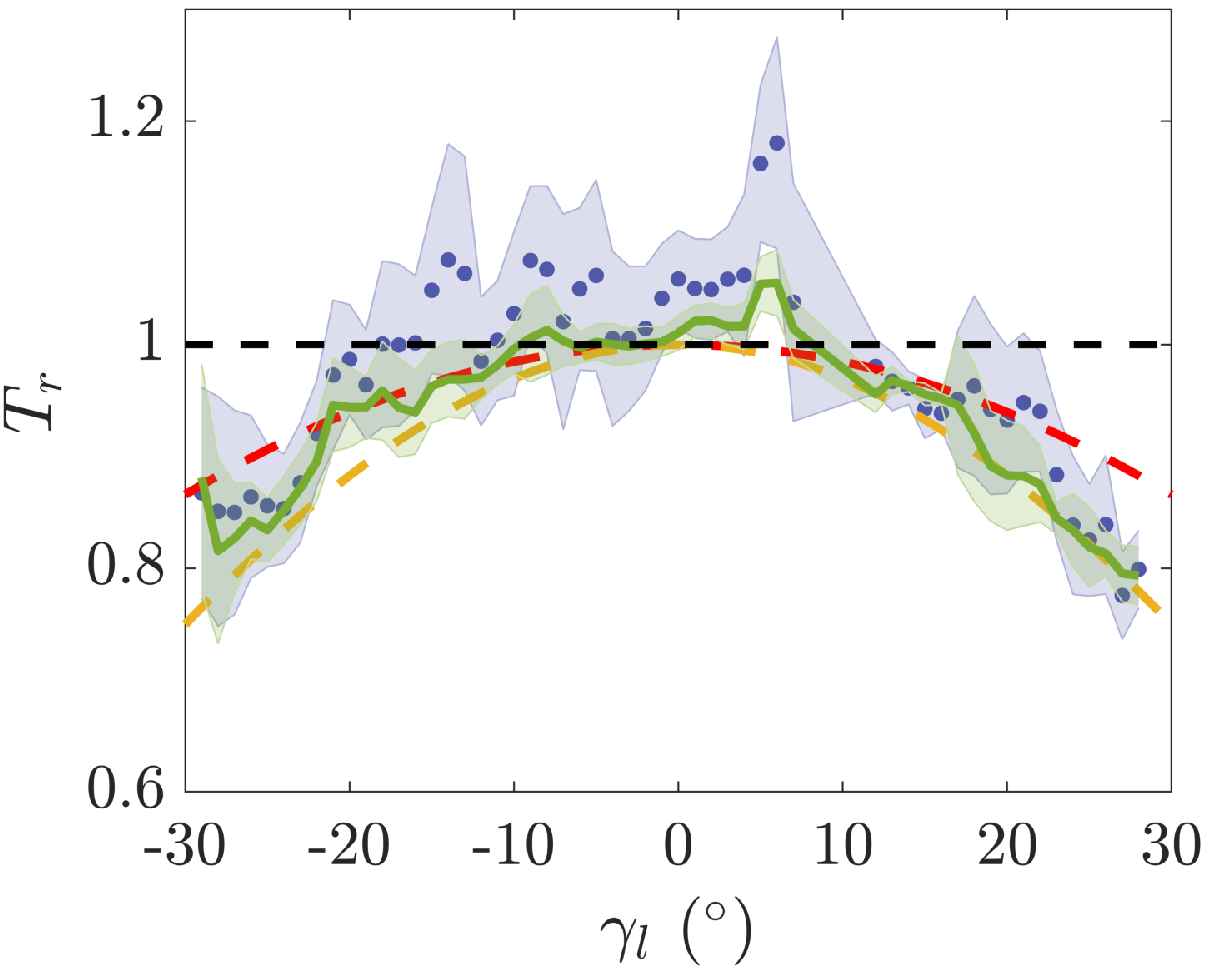} &
    \subfigimgthree[width=\linewidth,valign=t]{(c)}{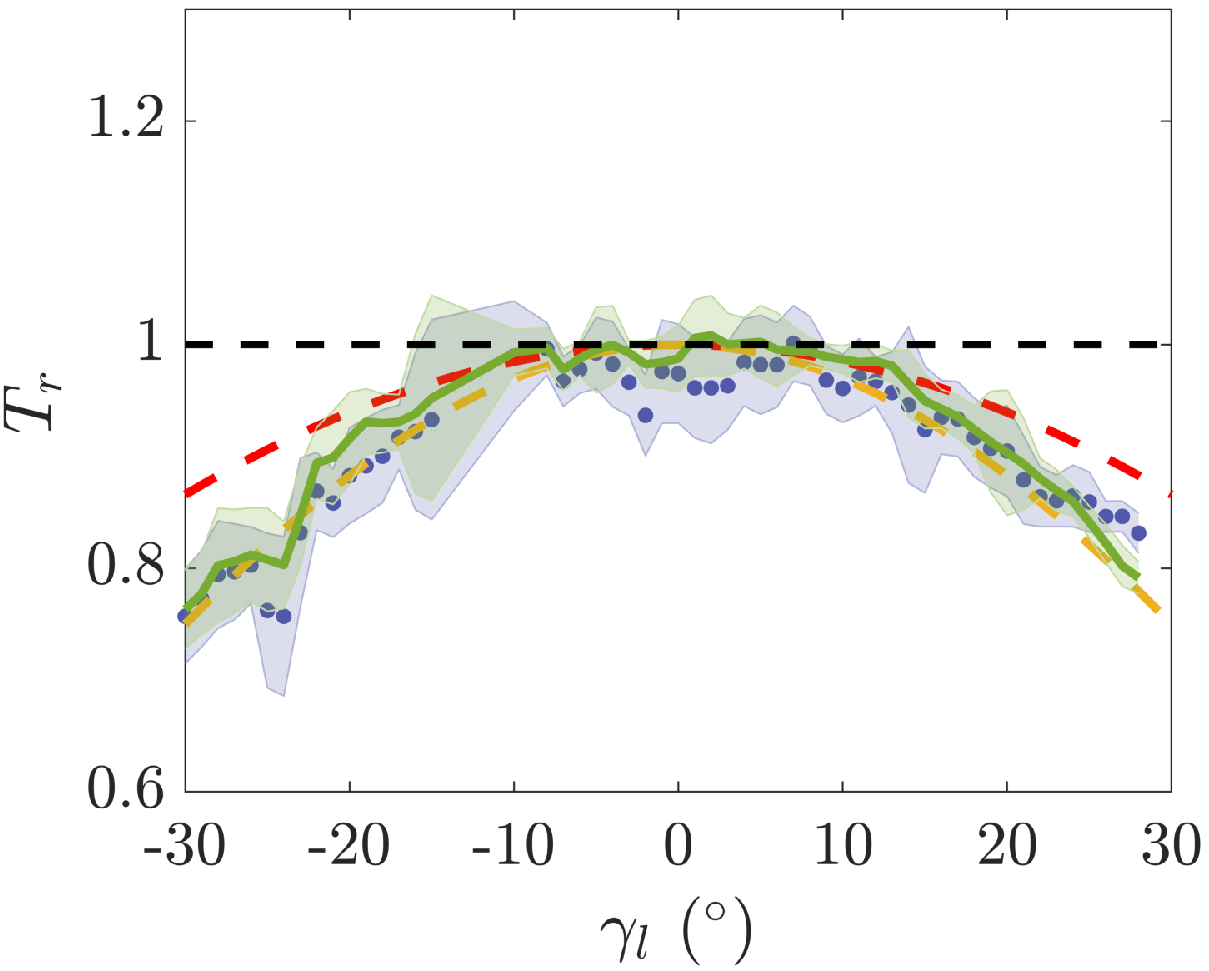}
    \end{tabular}
  \caption{Semi-empirical blade element model comparison. Torque ratio $T_r = T_{A1}/T_{A2}$ for incident wind directions of $-30^\circ<\alpha(z=z_h)<45^\circ$ as measured by the LiDAR. 
  The yaw misalignment values are binned in $1^\circ$ increments.
  Within each yaw misalignment bin, $10\%$ tails on the upper and lower ends of the $T_r$ PDF are removed.
  $\overline{T_r}[10\%-90\%]$ denotes the median of the central $80\%$ of the $T_r$ PDF.
  The yaw misalignment $\gamma_l$ is calculated by the LiDAR $\gamma_l= \alpha_{\mathrm{LiDAR}}-\beta_{A1}$.
  (a) All conditions of shear and veer are considered and the turbulence intensity is constrained between $0<TI<10\%$. 
  The number of resulting data points is $n=8376$.
  (b) $\alpha_v>0.2$, $\Delta \alpha_v>20$, $n=873$
  (c) $\alpha_v<0$, $\Delta \alpha_v>20$, $n=996$.
  Conditional bins with more than $5$ data points are shown.
        }
    \label{fig:Tr_empirical}
\end{figure}

\begin{figure}
  \centering
  \begin{tabular}{@{}p{0.33\linewidth}@{\quad}p{0.33\linewidth}@{\quad}p{0.33\linewidth}@{}}
    \subfigimgthree[width=\linewidth,valign=t]{(a)}{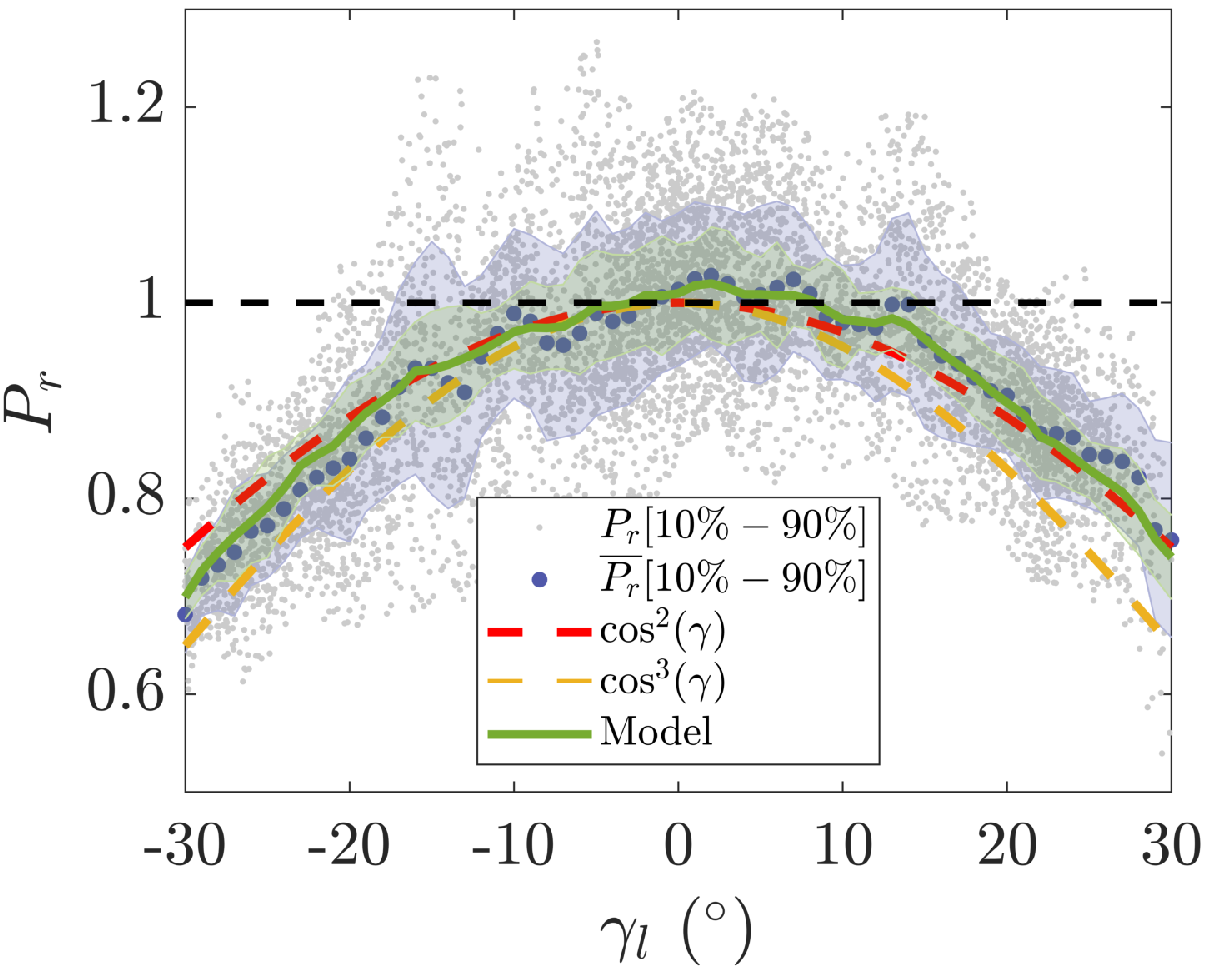} &
    \subfigimgthree[width=\linewidth,valign=t]{(b)}{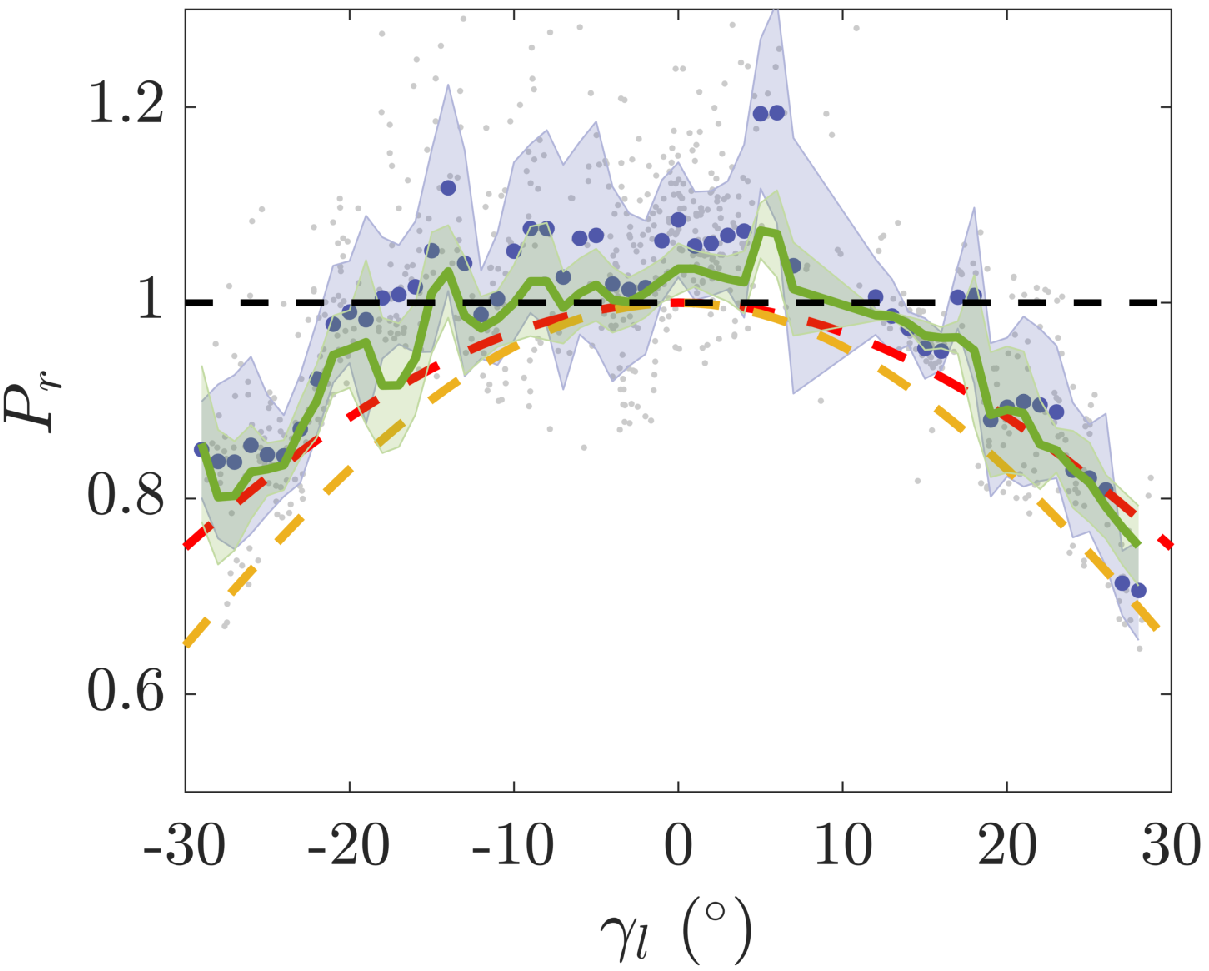} &
    \subfigimgthree[width=\linewidth,valign=t]{(c)}{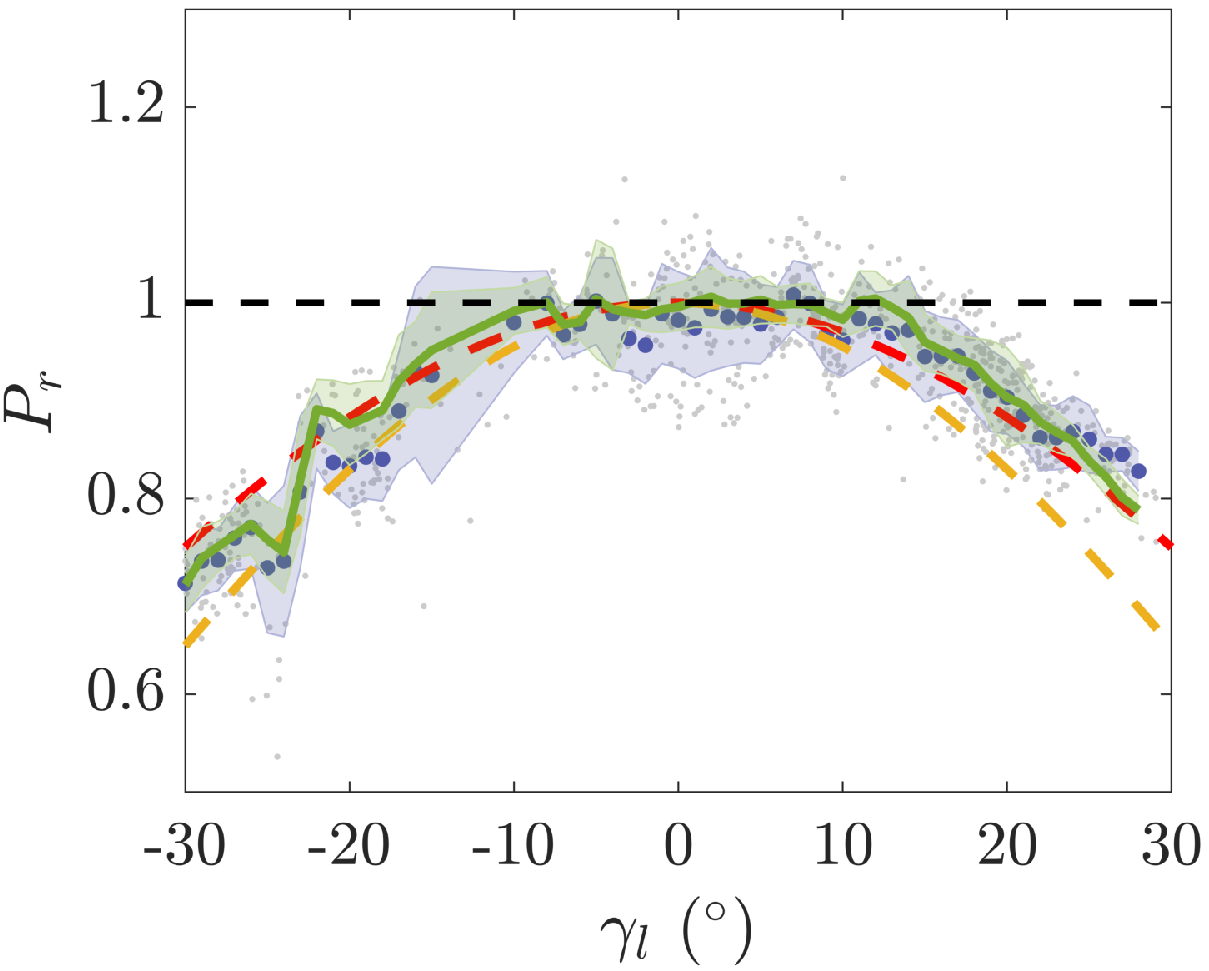}
    \end{tabular}
  \caption{Semi-empirical blade element model comparison. Power ratio $P_r = P_{A1}/P_{A2}$ for incident wind directions of $-30^\circ<\alpha(z=z_h)<45^\circ$ as measured by the LiDAR. 
  The yaw misalignment values are binned in $1^\circ$ increments.
  Within each yaw misalignment bin, $10\%$ tails on the upper and lower ends of the $P_r$ PDF are removed.
  $\overline{P_r}[10\%-90\%]$ denotes the median of the central $80\%$ of the $P_r$ PDF.
  The yaw misalignment $\gamma_l$ is calculated by the LiDAR $\gamma_l= \alpha_{\mathrm{LiDAR}}-\beta_{A1}$.
  (a) All conditions of shear and veer are considered and the turbulence intensity is constrained between $0<TI<10\%$. 
  The number of resulting data points is $n=8376$.
  (b) $\alpha_v>0.2$, $\Delta \alpha_v>20$, $n=873$
  (c) $\alpha_v<0$, $\Delta \alpha_v>20$, $n=996$.
  Conditional bins with more than $5$ data points are shown.
        }
    \label{fig:pr_full_data_empirical}
\end{figure}

There are a few potential sources of discrepancy between the model presented in \S \ref{sec:model_quant} and the field experiment data.
There is uncertainty associated with the impact of yaw misalignment on the measurements of the wind turbine nacelle-mounted wind speed and direction sensors \cite{pedersen2004wind}.
These measurements in turn dictate the turbine control system operational state.
Further, there is uncertainty associated with the wind direction calibrations (such that $0^\circ$ corresponds to true north) for the yaw actuating and yaw aligned turbines, as well as the profiling LiDAR.
This uncertainty is estimated to be approximately $\pm 1^\circ$ for each device.
For the model, higher order aeroelastic effects on the blades may modify the incident angle of attack $\phi-\psi$ in Eqs. \ref{eq:w} and \ref{eq:phi}, which could correspondingly modify the solution for $\Omega(\gamma)$.
Variations in axial induction over the rotor area were not considered, as the applicability of these empirical corrections for sheared and veered conditions is uncertain, and could be examined in future work. 
The quasi-static model assumes that the aerodynamic and generator torques are in equilibrium, which may not always hold for a given one-minute average due to the underlying dynamics of the generator torque control system.
Finally, the blade element model captures one-minute averaged variations in wind speeds, but higher frequency or intermittent incident wind content, such as wind gusts, which have a nonlinear impact on the power, torque, and angular velocity were not considered.
The angular velocity ratios $\Omega_r=\Omega(\gamma)/\Omega(\gamma=0)$ for the three wind conditions are shown in Figure \ref{fig:omega}.
While the model is able to predict the qualitative trends of $\Omega(\gamma)$, there are quantitative discrepancies, especially for $\gamma>0$.
In order to alleviate these issues while maintaining the analytic nature of the model presented in \S \ref{sec:model_quant}, we perform a semi-empirical model calculation where the resulting value of $\Omega_\gamma$ is used to predict $T_r$ and $P_r$.
While this method will not be available in a practical application setting since it requires a field experiment to measure $\Omega_\gamma$, this will serve as a validation of the model for the prediction of $T_r$ and $P_r$, which are not a trivial result of $\Omega_\gamma$ (see $T_r$ and $P_r$ derivation in \S \ref{sec:model_quant}).
The model results for $T_r$ and $P_r$ are given in Figures \ref{fig:Tr_empirical} and \ref{fig:pr_full_data_empirical}, respectively.

\begin{figure}
    \centering
    \includegraphics[width=0.75\linewidth]{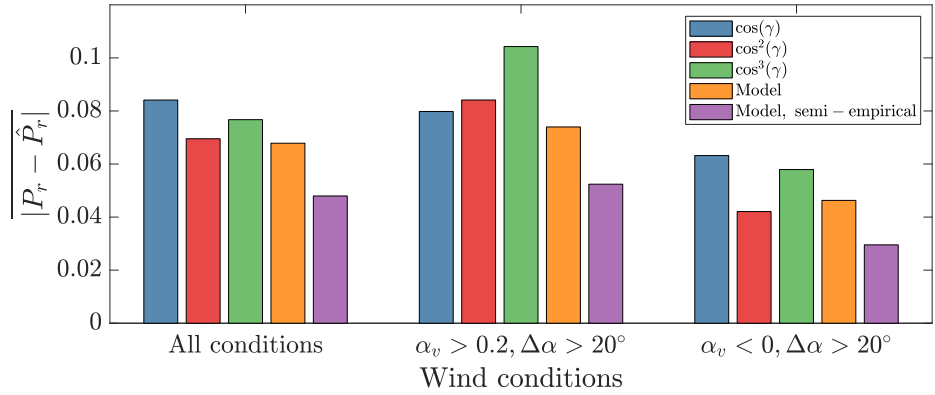}
    \caption{Mean absolute error between the measured power ratio $P_r$ and the predicted power ratio $\hat{P_r}$ for various cosine models, the predictive physics-based blade element model, and the semi-empirical blade element model where the model is provided $\Omega_r$.
    }
    \label{fig:mae}
\end{figure}

The mean absolute errors of the predicted $P_r$ associated with various cosine models, the physics-based blade element model, and the semi-empirical blade element model are shown in Figure \ref{fig:mae}.
The semi-empirical model has the lowest mean absolute error for all wind condition cases.
The $\cos^2(\gamma_l)$ achieves the lowest error of the cosine models except for $\alpha_v>0.2$ data.
The physics-based model has lower error than all cosine model approximations for all cases except for the $\cos^2(\gamma_l)$ model for the negative shear dataset, highlighting the asymmetric, complex influence of the incident wind conditions.
It is worth noting that there is not a precise physical justification for the form of the cosine model (Eq. \ref{eq:pr_general}) or the associated value of the $P_p$ exponential factor (see e.g. discussion by Pederson (2004)\cite{pedersen2004wind} or Bastankhah \& Port{\'e}-Agel (2017) \cite{bastankhah2017wind}), and therefore the application of the correct $P_p$ to reduce the power ratio prediction error is unknown {\it a priori}, while the physics-based blade element model is fully predictive.
The quantitative agreement between the field data and the model presented in \S \ref{sec:model_quant} are significantly improved in the semi-empirical formulation, with the model capturing sharp, nonmonotonic trends present in the field data with reasonable accuracy.
The success of the aerodynamic model presented in \S \ref{sec:model_quant} for qualitative predictions of $P_r$ without $\Omega_\gamma$ and improved quantitative predictions with $\Omega_\gamma$ suggest that the model can be used before wake steering control to estimate $P_r$ given the aerodynamic properties of the turbine of interest.

\begin{figure}
  \centering
  \begin{tabular}{@{}p{0.33\linewidth}@{\quad}p{0.33\linewidth}@{\quad}p{0.33\linewidth}@{}}
    \subfigimgthree[width=\linewidth,valign=t]{(a)}{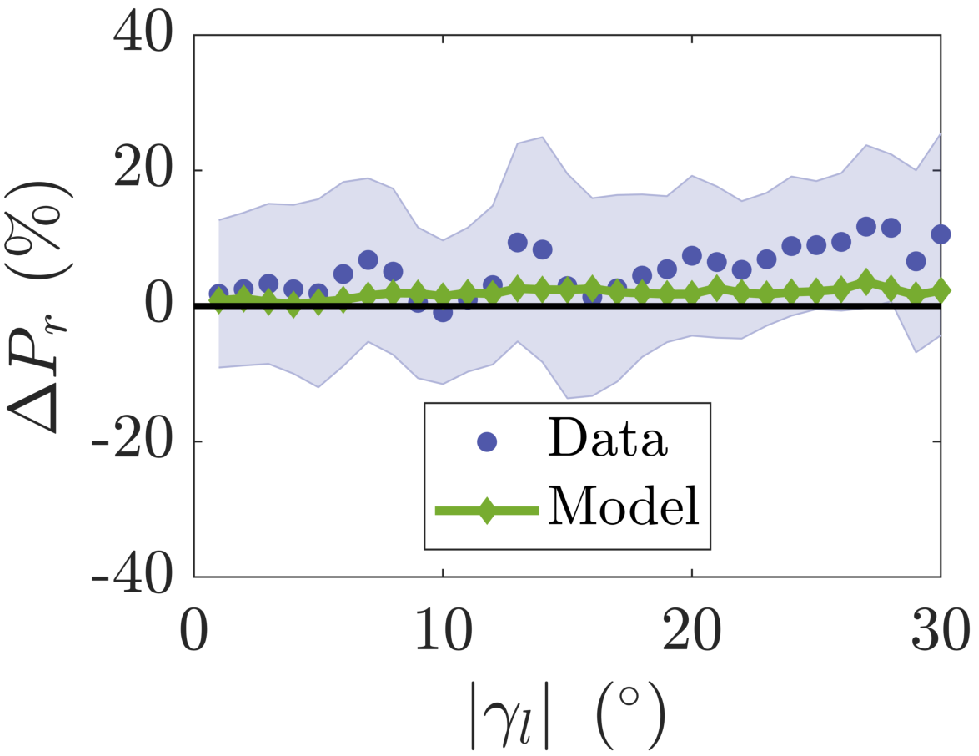} &
    \subfigimgthree[width=\linewidth,valign=t]{(b)}{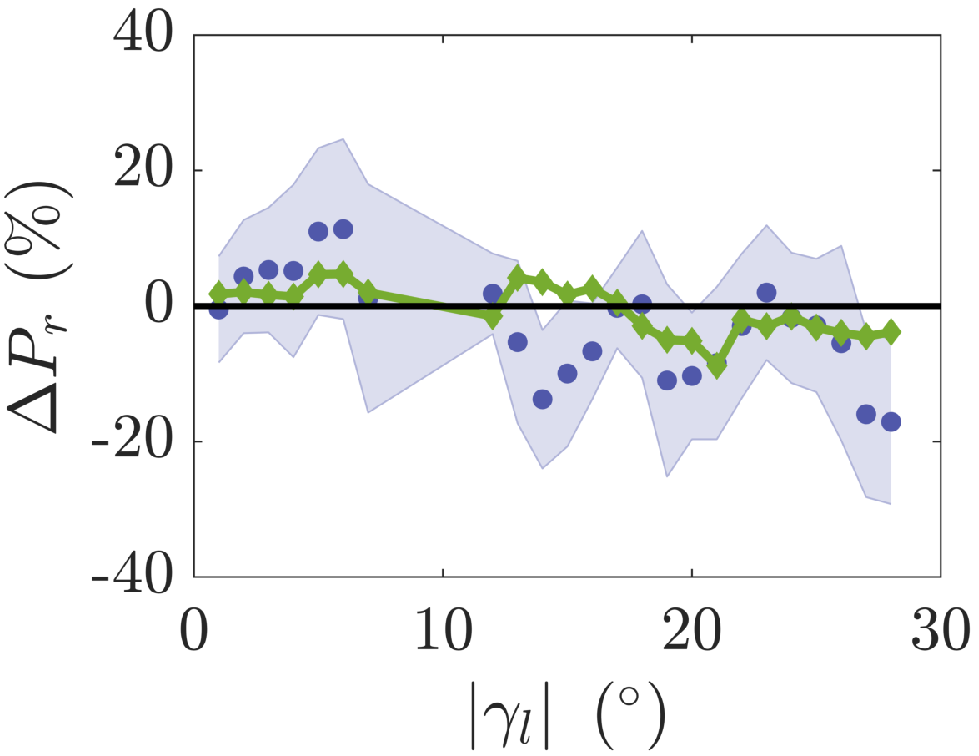} &
    \subfigimgthree[width=\linewidth,valign=t]{(c)}{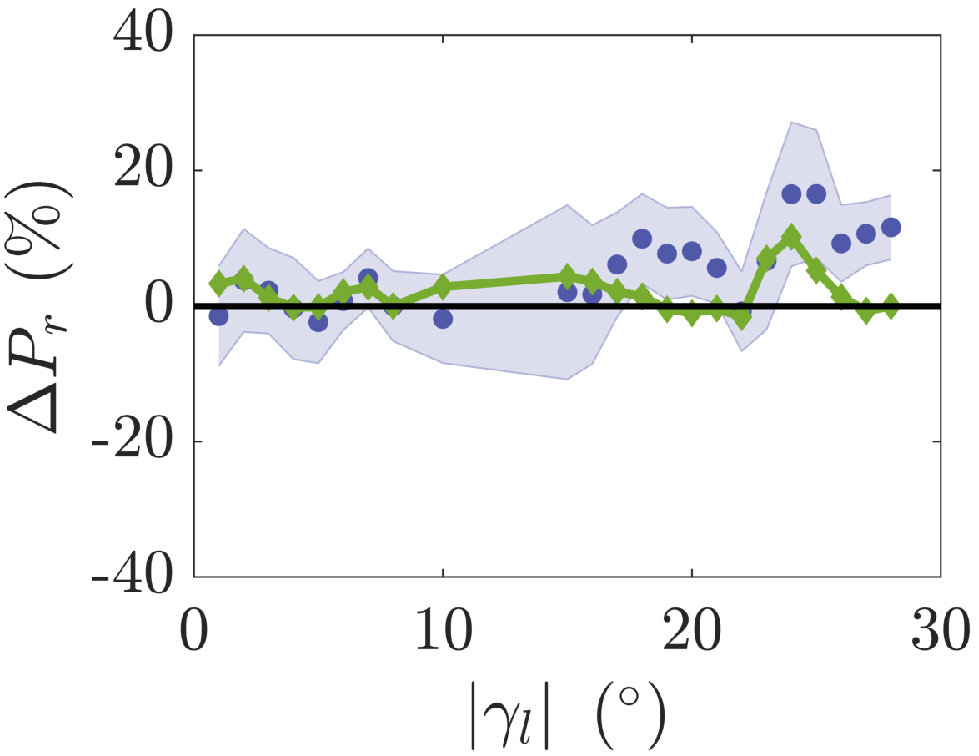}
  \end{tabular}
  \caption{Physics only model. The difference in the power ratio between positive yaw misalignment ($P_{r+}$) and negative yaw misalignment ($P_{r-}$) with a fixed absolute value computed as $\Delta P_r =  2(P_{r+} - P_{r-})/(P_{r+} + P_{r-})$ for $\Delta \alpha>20$ and (a) all shear cases, (b) $\alpha_v>0.2$ and (c) $\alpha_v<0$.
  Conditional bins with more than $5$ data points are shown.
        }
    \label{fig:asymmetry_physics}
\end{figure}

In order to further detail the asymmetric trends of the measured $P_r(\gamma)$ and the model, the normalized difference between the power ratio for positive and negative yaw misalignment is computed
\begin{equation}
\Delta P_r = \frac{P_{r+} - P_{r-}}{\frac{1}{2}(P_{r+} + P_{r-})},
\label{eq:asymmetry}
\end{equation}
where $P_{r+}$ indicates $P_r(\gamma_l>0)$ and $P_{r-}$ indicates $P_r(\gamma_l<0)$.
For cosine models of the power ratio (Eq. \ref{eq:pr_general}), $\Delta P_r=0$ $\forall \gamma$ by definition.
The profiles of $\Delta Pr$ for $-\infty<\Delta \alpha<\infty$ and $-\infty<\alpha_v<\infty$, $\Delta \alpha>20^\circ$ and $\alpha_v>0.2$, and $\Delta \alpha>20^\circ$ and $\alpha_v<0$ are shown in Figure \ref{fig:asymmetry_physics} for the experimental data and the physics-based model.
The full dataset exhibits an asymmetry such that $P_r(\gamma_l>0)>Pr(\gamma_l<0)$.
The model predicts a slightly higher value for $P_r(\gamma_l>0)$ and is within one standard deviation of the experimental data but the quantitative agreement is not precise.
For restricted positive or negative values of $\alpha_v$, as shown in Figure \ref{fig:asymmetry_physics}(b,c), the model reproduces the qualitative trend observed in the field data as well as an improved quantitative accuracy.
Interestingly, there are occasional discrete modulations in $\Delta P_r(\gamma_l)$ that result in a nonmonotonic profile as a function of $\gamma_l$.
Since the model, in general, quantitatively captures these discrete events, there are two likely explanations for this nonmonotonic behavior which act in tandem.
Given the strong veering and shearing conditions observed during the experiment, the hub height yaw misalignment angle which produces maximum power is not necessarily zero, as also discussed by Kragh \& Hansen (2014) \cite{kragh2014load} with respect to shear and Murphy {\it et al.} (2019)\cite{murphy2019wind} with respect to shear and veer.
Therefore, the peak $P_r$ may not occur at $\gamma_l=0$.
Further, even with the wind condition filters on $\alpha_v$ and $\Delta \alpha$, a variety of wind conditions are realized due to the complex nature of the turbulent ABL flow in a field environment (see also randomly selected velocity profiles in Figure \ref{fig:shear}).
Given the variety of velocity and direction profiles realized within the wind condition bins, the trends in $P_r$ are not isolated to $\gamma_l$ but also have a functional dependence on the wind conditions themselves.
Since the model resolves the leading-order effects of these variations in $u(z)$ and $\alpha(z)$, the model captures these discrete events with reasonable accuracy.

\begin{figure}
  \centering
  \begin{tabular}{@{}p{0.33\linewidth}@{\quad}p{0.33\linewidth}@{\quad}p{0.33\linewidth}@{}}
    \subfigimgthree[width=\linewidth,valign=t]{(a)}{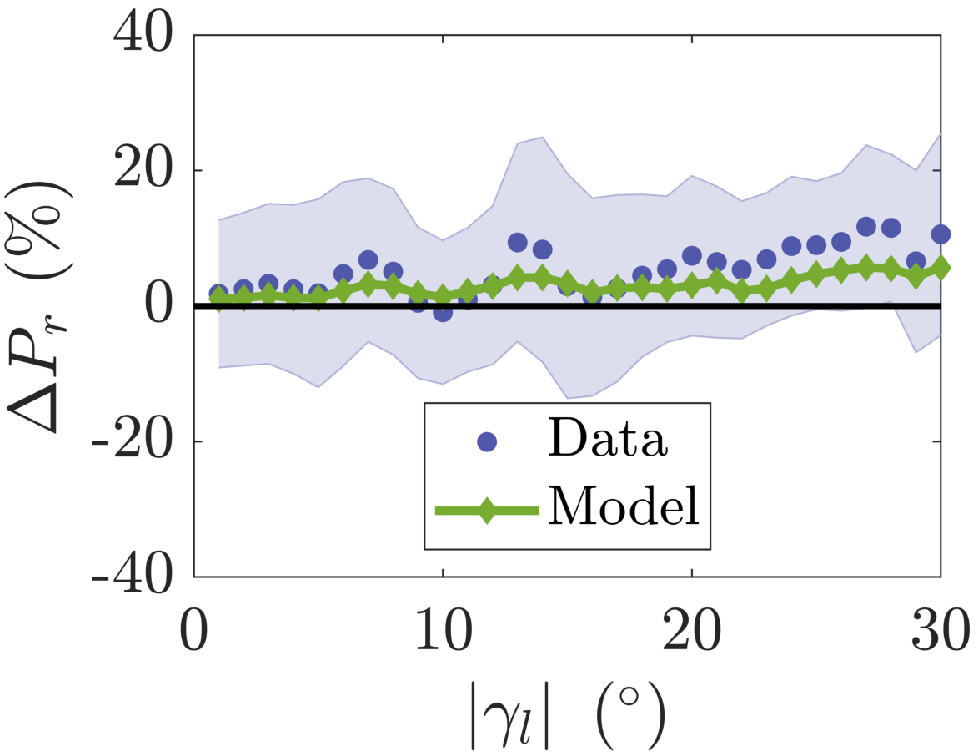} &
    \subfigimgthree[width=\linewidth,valign=t]{(b)}{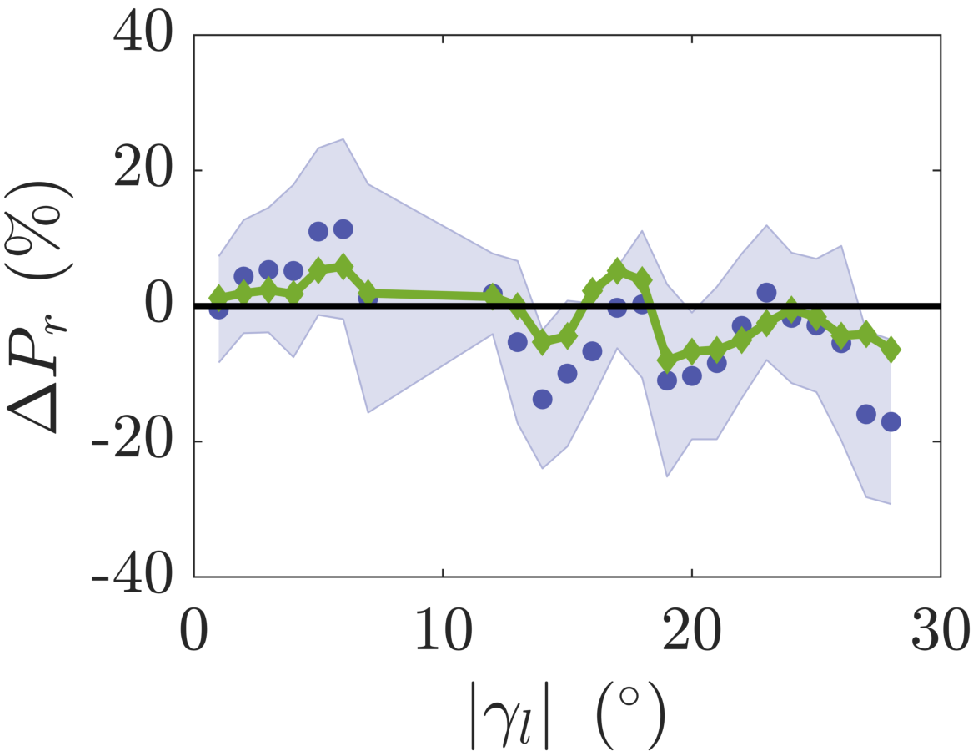} &
    \subfigimgthree[width=\linewidth,valign=t]{(c)}{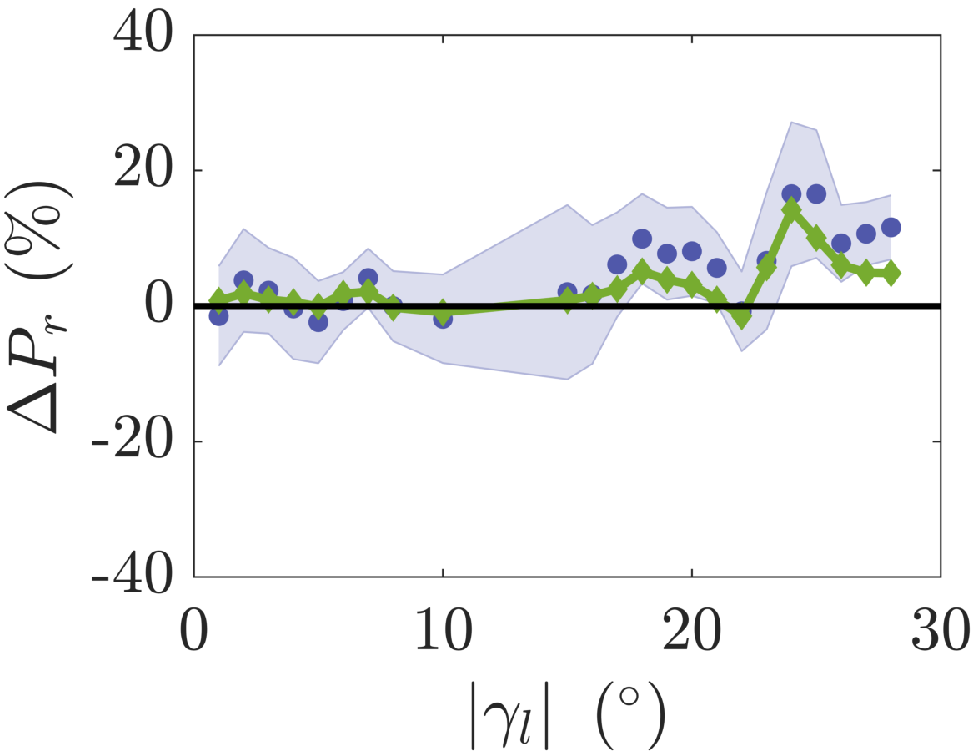}
  \end{tabular}
  \caption{Semi-empirical model. The difference in the power ratio between positive yaw misalignment ($P_{r+}$) and negative yaw misalignment ($P_{r-}$) with a fixed absolute value computed as $\Delta P_r =  2(P_{r+} - P_{r-})/(P_{r+} + P_{r-})$ for $\Delta \alpha>20$ and (a) all shear cases, (b) $\alpha_v>0.2$ and (c) $\alpha_v<0$.
  Conditional bins with more than $5$ data points are shown.
        }
    \label{fig:asymmetry_empirical}
\end{figure}

As with the $P_r$, we can also compute $\Delta P_r$ using the semi-empirical approach wherein the model is provided $\Omega_\gamma$.
The asymmetry of the power ratio $\Delta P_r$ for the semi-empirical model is shown in Figure \ref{fig:asymmetry_empirical}, where the qualitative and quantitative experimental results are reproduced within the errorbars of the field data for nearly all data-points.

\section{Discussion and implications for wake steering control}
\label{sec:discussion}

The full-scale field experimental results presented in \S \ref{sec:results} confirm the expectation that wind turbines in yaw misalignment will exhibit asymmetric power production as a function of the sign of the yaw misalignment angle depending on the incident wind conditions.
For wind velocity profiles that follow a power law with a positive shear exponent and exhibit clockwise Ekman turning associated with Coriolis forces in the northern hemisphere, negative yaw misalignment leads to enhanced power production for the yawed turbine compared to positive yaw misalignment.
On the other hand, for strongly stable conditions where positive veering and a sub-geostrophic jet emerge, positive yaw misalignment is beneficial compared to negative yaw misalignment.
The asymmetric influence of the wind conditions on the power production of a yaw misaligned turbine are represented with the model proposed in \S \ref{bem_model}.
While the quantitative value of $P_r$, and asymmetry of $P_r$ as a function of the direction of the yaw misalignment, will depend on the wind turbine control system and local wind conditions, the simple model proposed in \S \ref{bem_model} predicts $P_r(\gamma)$ with reasonable accuracy, suggesting that the model can be used to estimate the $P_r(\gamma)$ for wind turbines without requiring months long field experiments.
Further improvements in the predictive capabilities of the model are expected if the influence of the yaw misalignment on the measurements of the nacelle-mounted turbine sensors, and therefore the torque controller, are quantified.
The quantitative predictions of the power ratio may also be improved if full aeroelastic solvers which incorporate the effects of shear, veer, yaw misalignment, and turbine torque control are used.
However, the simple model proposed in \S \ref{sec:model_quant} captures the trends of the complex field data, highlighting the impact of the influence wind shear and veer on the power production of wind turbines in yaw misalignment. 

As discussed in \S \ref{sec:intro}, previous simulations and field experiments have shown a potential asymmetry in the power production of a wake steering scenario based on the direction of the yaw misalignment for fixed magnitudes of yaw.
The asymmetries in $P_r$, $T_r$, and $\Omega_r$ found in this field experiment and modeled in \S \ref{sec:model_quant} represents another potential cause of asymmetry in the wind turbine array power production given wake steering control as a function of the sign of $\gamma$, aside from the curled wake \cite[]{howland2016wake, fleming2018simulation}, Coriolis effects \cite[]{archer2019wake}, or the wake rotation direction \cite{gebraad2016wind}.
Importantly, the asymmetry in the success of wake steering as a function of yaw misalignment is case specific, the turbine array power production is not always higher given $\gamma>0$ than $\gamma<0$, as this asymmetry depends on the alignment of the wind turbines and the wind conditions.
The same magnitude and direction of asymmetry has not been observed in all studies.
Since the asymmetry in the power, torque, and angular velocity ratios of the upwind, yaw misaligned turbine depends on the wind conditions, the asymmetry in the total wind farm power given a wake steering strategy is also expected to depend on the characteristics, and in particular the stability, of the ABL.

When maximizing wind farm power production using wake steering, the optimal yaw misalignment angles, as well as the resulting power production increase, depend strongly on the power ratio.
Recent simulations have shown that for an incorrect estimate of $P_p$ in the simple power ratio model $P_r=\cos^{P_p}(\gamma)$, the power production for the wind farm can be reduced by wake steering compared to standard individual turbine control \cite{howland2020optimal}.
The results of this field experiment suggest that the standard, symmetric $P_p$ model is insufficient and will lead to asymmetric and site- and time-dependent errors in $P_r$. 
Instead, the site- and time-specific wind speed and direction profiles, measured using MET masts or LiDARs, should be leveraged to correct the $P_r$ model.
Future work should investigate the potential for ground-based extrapolation methods to provide the wind conditions (e.g. Lackner {\it et al.} (2010)\cite{lackner2010new}) in the absence of LiDAR or MET mast wind profile measurements.

While forces on the wind turbine were not measured in the field experiment, the asymmetric behavior of the power production is also expected in the axial force (Eq. \ref{eq:fa}).
Future work should investigate the joint influence of shear, veer, and yaw misalignment on the blade bending moments, which are influenced by yaw misalignment \cite{damiani2018assessment}.

\section{Conclusions}
\label{sec:conclusions}

A field experiment was performed at a wind farm in northwest India involving multiple utility-scale wind turbines. 
The power production of a freestream wind turbine in yaw is asymmetric depending on the direction of the yaw misalignment.
The asymmetry in the power as a function of yaw is chiefly caused by the incident wind speed and direction profiles, the direction of the wind turbine blade rotation, the turbine control system, and potential asymmetric effects on turbine sensor systems.
Therefore, for differing incident wind conditions during a typical diurnal atmospheric boundary layer evolution, the power production of a freestream turbine as a function of yaw, and its associated asymmetry, may be modified.

The angular velocity of a variable speed wind turbine which uses a generator torque control system does not follow $\cos(\gamma),$ and instead, depends jointly on the yaw misalignment and incident wind conditions.
The angular velocity $\Omega(\gamma)$ is a consequence of the generator torque control system and was persistently larger than $\Omega(\gamma=0) \cdot \cos(\gamma)$ for the yaw misaligned turbine of interest in this study. 

A model for the prediction of the power of a yaw misaligned turbine for arbitrary inflow wind conditions and turbine geometry was developed.
Previous model approaches predict that the power of a yaw misaligned turbine operating in freestream conditions $P(\gamma) \sim \cos^3(\gamma)$, which differs from experimental measurements.
The prediction of $P(\gamma) \sim \cos^3(\gamma)$ directly follows from an assumption that $\Omega(\gamma)=\Omega(\gamma=0)\cdot \cos(\gamma)$; this assumption was found to be inaccurate for a generator torque controlled variable speed turbine.
The current model, which calculates the angular velocity as a function of the generator torque control system, aerodynamic forces, yaw misalignment, and wind velocity and direction profiles, predicts that $P(\gamma) \approx P(\gamma=0) \cos^2(\gamma)$ for the presently studied turbine, with asymmetric deviations caused by the incident wind conditions.
It is important to note that the specific scaling predicted by the model, and achieved in practice by the wind turbine, will depend on the turbine generator torque controller and the incident wind conditions at the wind farm site, but in general the yaw misaligned power will not follow $\cos^3(\gamma)$ unless the angular velocity follows $\cos(\gamma)$.
This reaffirms the expectation that the power of a yaw misaligned turbine is turbine model specific but also site-specific.
Future wake steering applications can leverage the model presented in this study to compute an expected power ratio $P_r$ for a given wind turbine of interest before field or computational deployments.
The simple model proposed in \S \ref{sec:model_quant} can be used with arbitrary inflow profiles, and can be coupled with dynamic wake models to estimate the power ratio for yawed wind turbines operating in the wakes of upwind turbines.
Future work should investigate the optimal generator torque control strategy to minimize the power production degradation as a function of the yaw misalignment depending on the incident wind conditions.

\begin{acknowledgements}
M.F.H. is funded through a National Science Foundation Graduate Research Fellowship under Grant No. DGE-1656518 and a Stanford Graduate Fellowship.
The authors would like to thank Sanjiva Lele for thoughtful suggestions on the blade element model and Varun Sivaram for support throughout the study.
The data used in this study is confidential at the request of the wind farm operator.
\end{acknowledgements}

\FloatBarrier

\appendix

\FloatBarrier

\bibliography{main}

\end{document}